\documentclass[
 aip,
 jmp,%
 amsmath,amssymb,
author-year,%
]{revtex4-1}

\pdfoutput=1

\usepackage{graphicx,hyperref}
\usepackage{dcolumn}
\usepackage{bm}
\usepackage{subfigure,amsmath,subeqnarray,fancybox,multirow,hhline,amsmath}

\newcommand{\pdiff}[3][]{\dfrac{\partial^{#1} #2}{\partial {#3}^{#1}}}

\begin{document}

\preprint{AIP/123-QED}

\title{The Dynamics of Liquid Drops and their Interaction with Solids of Varying Wettabilities}

\author{J. E. Sprittles}
 \email{james.sprittles@gmail.com}
\affiliation{Mathematical Institute, University of Oxford, Oxford, OX1 3LB, U.K.}

\author{Y. D. Shikhmurzaev}%
 \email{yulii@for.mat.bham.ac.uk}
 \affiliation{School of Mathematics, University of Birmingham, Edgbaston, Birmingham, B15 2TT, U.K. }

\date{\today}

\begin{abstract}
Microdrop impact and spreading phenomena are explored as an interface formation process using a recently developed computational framework.  The accuracy of the results obtained from this framework for the simulation of high deformation free-surface flows is confirmed by a comparison with previous numerical studies for the large amplitude oscillations of free liquid drops.  Our code's ability to produce high resolution benchmark calculations for dynamic wetting flows is then demonstrated by simulating microdrop impact and spreading on surfaces of greatly differing wettability. The simulations allow one to see features of the process which go beyond the resolution available to experimental analysis. Strong interfacial effects which are observed at the microfluidic scale are then harnessed by designing surfaces of varying wettability that allow new methods of flow control to be developed.
\end{abstract}


\maketitle

\section{Introduction}

The impact and spreading of liquid drops on solid substrates is the key element of a range of technological processes. Examples include spray cooling of surfaces, crop spraying, spray coating, solder jetting and DNA synthesis \citep{grissom81,bergeron01,attinger00,maier00}. Thus far, research activity has been overwhelmingly devoted to the behaviour of millimetre-sized drops \citep{lesser83,rein93,yarin06} where the spatio-temporal scales of interest allow experiments to be performed routinely \citep{rioboo02}. However, increasingly, there is an interest in the dynamics of microdrops, whose behaviour is critical to the functioning of a number of microfluidic devices \citep{quake05}. The ink-jet printer is one such device which, as well as being utilized in the graphic arts, is beginning to become a viable alternative to traditional fabrication methods \citep{gao94,hong00,calvert01,burns03}, such as in the cost effective printing of P-OLED displays \citep{singh10} or the building of complex 3D structures through additive manufacturing \citep{derby10}. In such processes, the interaction of the microdrops with the solid substrate on which it impacts is directly related to the quality of the product, so that it is important to be able to predict and understand the behaviour of microdrops in such situations.

Much research, currently confined to the millimetre scale, has focused on how the wettability of a solid can dramatically influence a drop's dynamics \citep{renardy03} and how surfaces can be designed with areas of high and low wettability in order to control the spreading of drops on them and enhance the precision of the process to correct for inherent inaccuracies in droplet deposition \citep{mock05}.  Given the even larger surface-to-volume ratio of microdrops, so that surface effects become more dominant, the possibilities for flow control using a pre-patterned solid substrate are significant; however, thus far very few experimental or theoretical papers have addressed this promising new area.

The dynamics of microdrops as they impact on the solid substrate is difficult to observe experimentally \citep{dong07,vandam04}, especially with a sufficiently high temporal resolution, and characteristics such as the flow field inside the drop or stress acting on the substrate are almost impossible to measure.  Therefore, it becomes necessary to rely on a theory which, once validated against the data from experiments on millimetre-sized drops and other relevant free-surface flows, would allow one to obtain reliable information about this process.

So far, the main emphasis in research on microdrop spreading has been on the influence of additional physical effects on the drop's dynamics such as heat and mass transfer \citep{bhardwaj08}; polymeric properties of the liquid \citep{perelaer09}; entrapment of bubbles under the drop \citep{vandam04}; spreading on, and imbibition into, porous media \citep{clarke02,holman02}; evaporation \citep{lim09}; and solidification \citep{attinger00}.  The focus of our research programme is to begin by accurately capturing the process of dynamic wetting, which is the key physical effect in the drop impact phenomenon, and develop a benchmark numerical platform capable of incorporating complex mathematical models that describe the essential features of this process.  Once this aspect is resolved, additional physical factors such as heat transfer, interaction with external fields, etc, can be built into the framework as required.  An example of one such additional physical effect was given in \onlinecite{sprittles07,sprittles09}, where the model was extended to account for surfaces of inhomogeneous wettability.

The issues surrounding the modelling of dynamic wetting flows are well known and have been the subject of several reviews \citep[e.g.][]{dussan79,blake06} and monographs \citep[e.g.][]{degennes85,shik07}, with a general discussion of the merits of microscopic, mesoscopic and macroscopic modelling approaches given in \onlinecite{shik11} and the discussion notes that followed. Here, we focus on the self-consistent framework of continuum mechanics where, in particular, it has been established that the classical model of fluid mechanics must be modified to allow for a solution to exist \citep{huh71,dussan79}. A common way to achieve this is to use a `slip model' where the no-slip condition is modified to allow for slip between the liquid and the solid near the moving contact line, e.g.\ using the Navier condition \citep{navier23}, whilst the contact angle is prescribed as a function of the contact-line speed and material constants \citep[e.g.][]{greenspan78}. When incorporated into numerical software, such models have been shown to produce adequate results for the spreading of millimetre-sized drops at relatively low impact speeds, where experiments can be easily analyzed to allow for the development of a semi-empirical analysis of the phenomenon \citep{pasandideh96,yokoi09}.

An open question is whether the models that have been specifically developed for millimetre-sized drop dynamics can predict the behaviour of drops across a range of scales, i.e. towards micro/nanodrops, at a range of impact speeds.

A step in the direction of answering this question was taken in the studies of \onlinecite{bayer06} and \onlinecite{sikalo05}.  Their results show that, even with millimetre-sized drops, the contact angle is not simply a function of the contact-line speed, but is actually determined by the entire flow field. In other words, the dependence of the contact angle on the contact-line speed for given materials of the system is non-unique; it varies with the speed of impact, i.e. it depends on the particular flow. For an illustration of this effect we refer the reader to Figure~$14$ in \onlinecite{bayer06} and Figure~9 in \onlinecite{sikalo05}. This effect of the flow field on the contact angle is well known in the process of curtain coating where it has been termed the `hydrodynamic assist of dynamic wetting' \citep{blake94,clarke06}. Similar dependencies of the dynamic contact angle on the flow field have been noted in the spreading of a liquid between parallel plates \citep{ngan82}, the imbibition of liquid into capillaries \citep{sobolev00,sobolev01} and in the coating of fibres \citep{simpkins03}; but these flows are yet to undergo the level of scrutiny that the curtain coating process has where it has been shown using careful finite element simulations that the effect cannot be described using any interpretation of the conventional `slip' models \citep{wilson06}.  Put simply, all currently available computational software, which implement the `conventional' (i.e.\ slip) models, are unable to describe this key physical effect which has already been shown to be critical for the optimization of one industrial process and, as experiments suggest, is critical to the understanding of microdrop impact and spreading phenomena.

Currently, the only model able to predict the influence of the flow field on the contact angle is the interface formation model. This model is based on the simple physical idea that dynamic wetting, as the very name suggests, is the process in which a fresh liquid-solid interface (a newly `wetted' solid surface) is formed.  The model is described in detail in \onlinecite{shik07} and has already been shown to be in agreement with experimental data in a range of different physical phenomena \citep{shik93,shik97,shik97a,shik05b,shik05c,shik05a,sprittles07,sprittles09}.  Notably, recent benchmark simulations for the technologically relevant phenomenon of a liquid-gas meniscus propagating through a cylindrical capillary tube \citep{sprittles_jcp} confirm that by modelling dynamic wetting as an interface formation process, we are able to demonstrate the effect of the flow field on the contact angle so that the aforementioned flows exhibiting assist-like behaviour can now be studied using our computational tool.

A further advantage of the interface formation model over conventional approaches is that it is able to naturally account for the influence of variations in the wettability of the solid surface on the bulk flow. Specifically, even when there is no free surface present, the interface formation model predicts that a single \citep{sprittles07} or multiple \citep{sprittles09} changes in the wettability of the solid substrate can disrupt a shear flow parallel to that solid.  The obtained results are in qualitative agreement with the predictions of molecular dynamics simulations \citep{priezjev05,qian05} and will become more important as the scale of the flow is reduced, i.e.\ as we consider micro and nanofluidic flows.

Computation of dynamic wetting flows is complex: besides the effects of capillarity, viscosity and inertia, one must also capture the physics of wetting which typically occurs on a length scale much smaller than that of the bulk flow.  As explained in detail in \onlinecite{sprittles11c}, the majority of publications in the field fail to accurately account for the wetting dynamics on the smaller length scale and, consequently, make it impossible to distinguish physical effects from numerical errors in their results. Such codes may provide realistically-looking results for millimetre-sized drops, where the accurate computation of the bulk dynamics may be sufficient, but on the micro/nano scale, where an increasing surface-to-volume ratio means that surface effects become more important, such codes are unreliable as the free surface shape near the contact line, specified by the contact angle, is not accurately determined.

The first steps in the development of a benchmark code for such flows was described in great detail in \onlinecite{sprittles11c}, where a user-friendly step-by-step guide is provided to allow the reader to reproduce all results. This framework was extended in \onlinecite{sprittles_jcp} to allow for the simulation of dynamic wetting as an interface formation process, with the approach robustly tested by checking its convergence both under mesh refinement, allowing practical recommendations on mesh design to be made, and, in limiting cases, to analytic asymptotic results. Furthermore, the developed computational tool was used to predict new physical effects, such as a new type of dependence of the dynamic contact angle on the flow geometry, and was seen to describe experimental data for the imbibition of a liquid into a capillary tube exceptionally well and significantly better than the Lucas-Washburn approach \citep{washburn21}. This situation is in complete contrast to commercial software which have not been validated as thoroughly for this class of flows and have been shown to converge to the wrong solution for similar flows \citep{hysing07}.

In contrast to the flows considered in \onlinecite{sprittles11c,sprittles_jcp}, impacting microdrops undergo extreme changes in shape.  Therefore, having outlined the model in Sec.~\ref{model} and the numerical approach in Sec.~\ref{numerics}, in Sec.~\ref{osc} to verify the code's accuracy for large changes in free surface shape we compare our calculations to a known test-case from the literature of oscillating liquid drops, where reliable results exist for an unsteady problem in which inertia, capillarity and viscosity are all prominent; it is also a problem of significant interest in its own right.

Having verified the code's ability to accurately simulate the dynamics of liquid drops, in Sec.~\ref{drops} we demonstrate its capabilities for drop impact and spreading phenomena. Here, we focus on the main physical effects and, in particular, the influence of a substrate's wettability on an impacting drop's behaviour as well as the ability of our computational tool to extract information that is hidden to experimental analysis and which allows new paths of enquiry to be proposed. Most promisingly, in Sec.~\ref{patt_surf} a novel mechanism of flow control is developed, based on pre-patterning a substrate with regions of high and low wettability, that allows the final shape of the drop to be controlled using only slight alterations in the impact speed.

\section{Modelling of Dynamic Wetting Phenomena}\label{model}

Consider the flow of an incompressible Newtonian liquid, of constant density $\rho$ and viscosity $\mu$, surrounded by a dynamically passive gas of a constant pressure $p_g$, so that the continuity and momentum balance equations are given by
\begin{equation}\label{ns}
\nabla\cdot\mathbf{u} = 0,\qquad \rho~\left[\pdiff{\mathbf{u}}{t} + \mathbf{u}\cdot\nabla\mathbf{u}\right] = -\nabla p + \mu\nabla^2\mathbf{u} + \rho \mathbf{g},
\end{equation}
where $t$ is time, $\mathbf{u}$ and $p$ are the liquid's velocity and pressure, and $\mathbf{g}$ is the gravitational force density.

The interface formation equations, which act as boundary conditions for the bulk equations (\ref{ns}), and have been described in great detail in \onlinecite{shik07}, are here simply listed with a very brief description given below. These equations consider interfaces as `surface phases' characterized by the surface density $\rho^s$, surface velocity $\mathbf{v}^s$ with which the density is transported and the surface tension $\sigma$ which can be viewed as a `two-dimensional pressure' taken with the opposite sign. On a liquid-gas free surface, we have
\begin{equation}\label{kin}
\pdiff{f}{t} + \mathbf{v}^s_1\cdot\nabla f = 0,
\end{equation}
\begin{equation}\label{stress} 	
\mu\mathbf{n}\cdot\left[\nabla\mathbf{u}+\left(\nabla\mathbf{u}\right)^T\right]\cdot
\left(\mathbf{I}-\mathbf{n}\mathbf{n}\right) +\nabla\sigma_1 =\mathbf{0},\qquad
p_g-p+ \mu\mathbf{n}\cdot\left[\nabla\mathbf{u}+\left(\nabla\mathbf{u}\right)^T\right]\cdot
\mathbf{n}=\sigma_1\nabla\cdot\mathbf{n},
\end{equation}
\begin{equation}\label{mass1}
\rho\left(\mathbf{u}- \mathbf{v}^s_{1}\right)\cdot\mathbf{n} = \left(\rho^{s}_1-\rho^{s}_{1e}\right)\tau^{-1},\qquad \pdiff{\rho^{s}_{1}}{t} + \nabla\cdot\left(\rho^{s}_{1}\mathbf{v}^{s}_{1}\right) = - \left(\rho^{s}_{1}-\rho^{s}_{1e}\right)\tau^{-1},
\end{equation}
\begin{equation}\label{others1}
4\beta\left(\mathbf{v}^{s}_{1||}-\mathbf{u}_{||}\right)=\left(1+4\alpha\beta\right)\nabla\sigma_1, \qquad\sigma_{1}=\gamma(\rho^s_{(0)}-\rho^{s}_{1}),
\end{equation}
where the a-priori unknown free surface is $f(\mathbf{r},t)=0$, with the inward normal $\mathbf{n} = \frac{\nabla f}{|\nabla f|}$,  the metric tensor of the coordinate system is $\mathbf{I}$ and the subscript $||$ denotes components parallel to the surface, which are obtained by convolution with the tensor $(\mathbf{I}-\mathbf{n}\mathbf{n})$.  Subscripts $1,2$ refer to variables on the free surface and liquid-solid interface, respectively. At a stationary liquid-solid interface, the equations of interface formation have the form
\begin{equation}\label{normal}
\mathbf{v}^s_2\cdot\mathbf{n}=0,\qquad
\mu\mathbf{n}\cdot\left[\nabla\mathbf{u}+\left(\nabla\mathbf{u}\right)^T\right]\cdot(\mathbf{I}-\mathbf{n}\mathbf{n}) + \hbox{$\frac{1}{2}$}\nabla
\sigma^s_{2}=\beta\mathbf{u}_{||},
\end{equation}
\begin{equation}\label{mass2}
\rho\left(\mathbf{u}-\mathbf{v}^s_2\right)\cdot\mathbf{n} = \left(\rho^{s}_{2}-\rho^{s}_{2e}\right)\tau^{-1},\qquad \pdiff{\rho^{s}_{2}}{t} + \nabla\cdot\left(\rho^{s}_{2} \mathbf{v}^{s}_{2}\right) = -\left(\rho^{s}_{2}-\rho^{s}_{2e}\right)\tau^{-1},
\end{equation}
\begin{equation}\label{others2}
\mathbf{v}^{s}_{2||}-\hbox{$\frac{1}{2}$}\mathbf{u}_{||}=\alpha\nabla\sigma_{2},\qquad \sigma_{2}=\gamma(\rho^s_{(0)}-\rho^{s}_{2}).
\end{equation}

Boundary conditions (\ref{kin})--(\ref{others2}) are themselves differential equations along the interfaces and are in need of boundary conditions at the contact line.  At a contact line where a free surface meets a solid, we have continuity of surface mass flux and the Young equation \citep{young05}, which balances the tangential components of the forces due to surface tensions acting on the contact line and hence determines the dynamic contact angle $\theta_d$:
\begin{equation}\label{cl}
\rho^s_1\left(\mathbf{v}^s_{1||} - \mathbf{U}_{c}\right)\cdot\mathbf{m}_1 + \rho^s_2\left(\mathbf{v}^s_{2||}  - \mathbf{U}_{c}\right)\cdot\mathbf{m}_2=0
, \qquad \sigma_{2}+\sigma_{1}\cos\theta_d=0.
\end{equation}
Here $\mathbf{m}_{i}$ are the unit vectors normal to the contact line and inwardly tangential to surface $i=1,2$, the velocity of the contact line is $\mathbf{U}_c$ and the surface tension of the solid-gas interface is assumed to be negligible.

On an axis of symmetry, with normal vector $\mathbf{n}_a$, the usual conditions of impermeability and zero tangential stress are applied
\begin{equation} \mathbf{u}\cdot\mathbf{n}_a=0, \qquad \mathbf{n}_a\cdot\left[\nabla\mathbf{u}+\left(\nabla\mathbf{u}\right)^T\right]\cdot(\mathbf{I}-\mathbf{n}_a\mathbf{n}_a)=\mathbf{0}, \end{equation}
together with conditions on the smoothness of the free surface where it meets such an axis $\mathbf{n}\cdot\mathbf{n}_a=0$ and the absence of a surface mass source/sink for the interface formation equations $\mathbf{v}^s_{||}\cdot\mathbf{n}_a=0$.

On the free surface, the standard kinematic equation is (\ref{kin}) and the equations balancing the tangential and normal stress acting on the interface from the liquid and gas with the capillary pressure are (\ref{stress}). On the liquid-solid interface, equations (\ref{normal}) state that the solid is impermeable and that the tangential component of the bulk velocity satisfies a generalized Navier condition which shows that slip on the liquid-facing side of the liquid-solid interface can be generated by both tangential stress \emph{on} the interface and variations of surface tension \emph{in} the interface. The model takes into account the mass exchange between the bulk and surface phases, in (\ref{mass1}) and (\ref{mass2}), that are associated with the relaxation of an interface with surface density $\rho^s$ towards its equilibrium state $\rho^s=\rho^s_e$ on characteristic relaxation time $\tau$.  The first equation in (\ref{others1}) and (\ref{others2}) shows that the tangential components $\mathbf{v}^s_{||}$ of the surface velocity are driven both by the bulk motion of the fluid and by gradients in surface tension along the interface.  The second equation in (\ref{others1}) and (\ref{others2}) is the surface equation of state, which here is taken in its simplest linear form, that determines the surface tension along the interface from the surface density and hence allows one to find the contact angle from the Young equation in (\ref{cl}). Therefore, the mechanism by which assist occurs becomes clear, with the flow able to influence the value of the surface tensions at the contact line and hence alter the contact angle. Estimates for the phenomenological material constants $\alpha$, $\beta$, $\gamma$, $\tau$ and $\rho_{(0)}^{s}$ have been obtained by comparing the theory to experiments in dynamic wetting, e.g.\ in \onlinecite{blake02}.

Notably, the wettability of the solid substrate is controlled by the equilibrium surface density on it $\rho^s_{2e}$, which is directly related to the surface tension in the liquid-solid interface via the equation of state (\ref{others2}) and hence to the equilibrium contact angle $\theta_e$ via the Young equation (\ref{cl}).  Then, as shown in \onlinecite{sprittles07}, variations in the wettability of the substrate, considered in Sec.~\ref{patt_surf}, can easily be built into the model by prescribing a spatial dependence for $\rho^s_{2e}$.

\section{A Computational Framework for Dynamic Wetting Phenomena}\label{numerics}

A framework for the simulation of dynamic wetting flows as an interface formation process has been developed in \onlinecite{sprittles_jcp}, which builds upon previous work \citep{sprittles11c} where the approach was formulated for the mathematically less complex conventional models.  These papers provide a step-by-step guide to the development of the code, curves for benchmark calculations and a demonstration of our platform's capabilities, by showing excellent agreement with experiments on capillary rise, so that, here, we shall just recapitulate the main details.

The CFD code is based on the finite element method and uses an arbitrary Lagrangian-Eulerian mesh design \citep{kistler83,heil04,wilson06} to allow the free surface to be accurately represented whilst bulk nodes remain free to move (snapshots of the mesh can be seen below in Figure~\ref{F:rebound_mesh}). This mesh is based on the bipolar coordinate system and is graded so that exceptionally small elements are used near the contact line to accurately capture the physics of interface formation there (see \onlinecite{sprittles_jcp} for estimates on the smallest element size required for given parameters), whilst in the bulk of the liquid larger elements are adequate and ensure the resulting problem is computationally tractable.

The result of our spatial discretization is a system of non-linear  differential algebraic equations of index two \citep{lotstedt86} which are solved using the second-order backward differentiation formula, whose application to the Navier-Stokes equations is described in detail in \onlinecite{gresho2}, using a time step which automatically adapts during a simulation to capture the appropriate temporal scale at that instant.  For example, the time steps required in the initial stages after impact of the drop are small compared to those used in the later stages where the drop often gently oscillates around its equilibrium position on a much longer characteristic time scale.

The described CFD code has been thoroughly validated for dynamic wetting phenomena in which the changes in free surface shape are moderate, but is here applied to microdrop impact and spreading where huge changes in free surface shape can be observed.  Therefore, continuing in the spirit of our previous publications, the code is first validated for this class of flows which, additionally, allows us to demonstrate how easily the the framework can be adapted to handle a variety of free surface flows.

\section{Oscillating Drops: Validation of the Unsteady Code for Large Free-Surface Deformation}\label{osc}

Consider the free oscillation of liquid drops in a parameter regime that will allow us to test our results against known numerical solutions for a high deformation unsteady flow which exhibits the competing forces of capillarity, inertia and viscosity. For small oscillations analytic results exist \citep{rayleigh79}, but for arbitrary viscosity and deformation numerical methods are required. The test case considered is for the standard model, which can be easily obtained from the interface formation model's equations for the free surface (\ref{kin})--(\ref{others1}) by setting the ratio $\tau/T$ of interfacial relaxation time $\tau$ to bulk time scale $T$ to zero (see \onlinecite{sprittles_jcp} for details), so that the same code can be used. In this case, the flow is fully characterized by the Reynolds, Stokes and capillary numbers, which are, respectively,
\begin{equation}\label{nd1}
Re=\frac{\rho U L}{\mu},\qquad St = \frac{\rho g L^{2}}{\mu U},\qquad Ca=\frac{\mu U}{\sigma_{1e}},
\end{equation}
where $L$ and $U$ are the scales for length and velocity, and $\sigma_{1e}$ is the equilibrium surface tension on the free surface.  Here, we consider oscillation in zero-gravity so that $St=0$.

\begin{figure}
     \centering
     \begin{minipage}[c]{.45\textwidth}     \centering
     \subfigure[t=0]   {\includegraphics[viewport=-65  0  235 300,angle=0,scale=0.42]{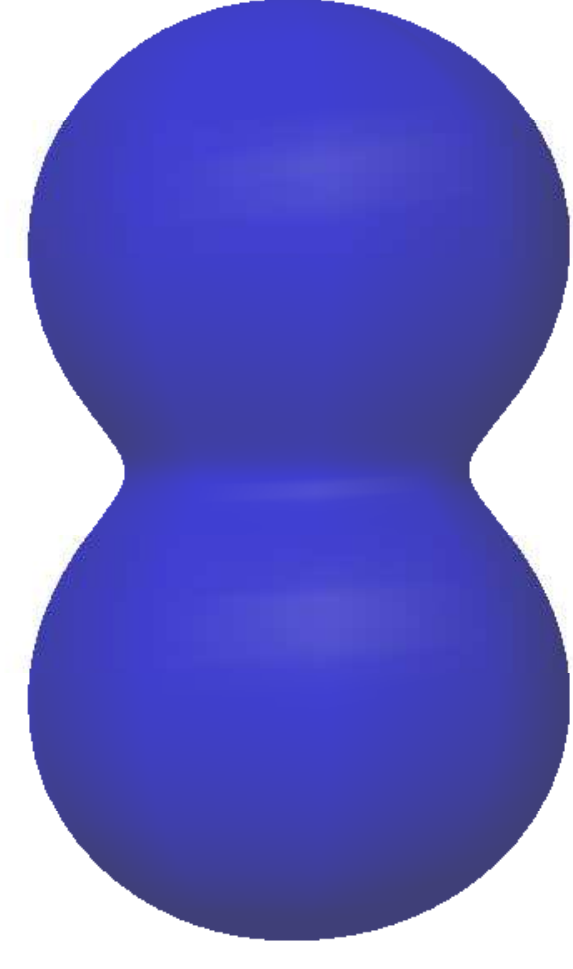}}
     \subfigure[t=0.42]{\includegraphics[viewport=-65  0  235 300,angle=0,scale=0.42]{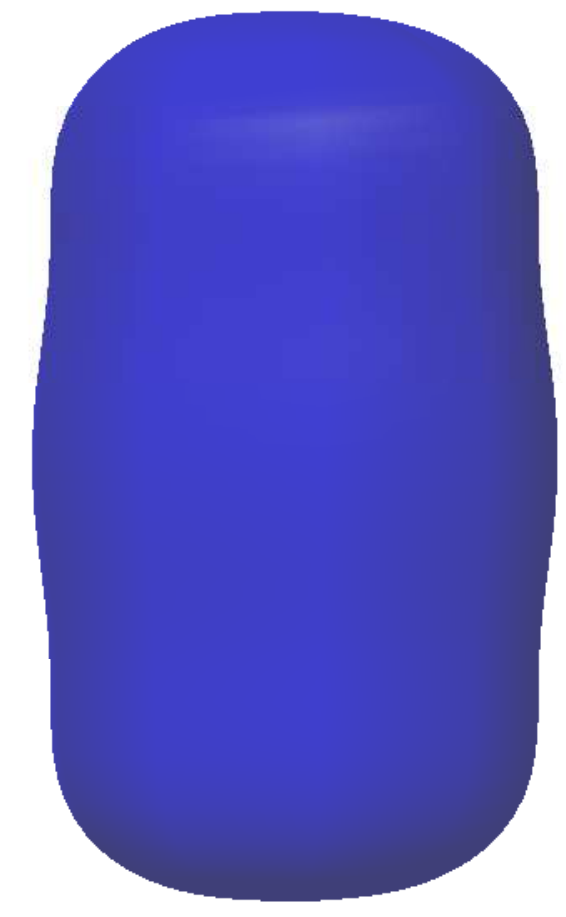}}
     \subfigure[t=0.84]{\includegraphics[viewport=-50  0  250 300,angle=0,scale=0.42]{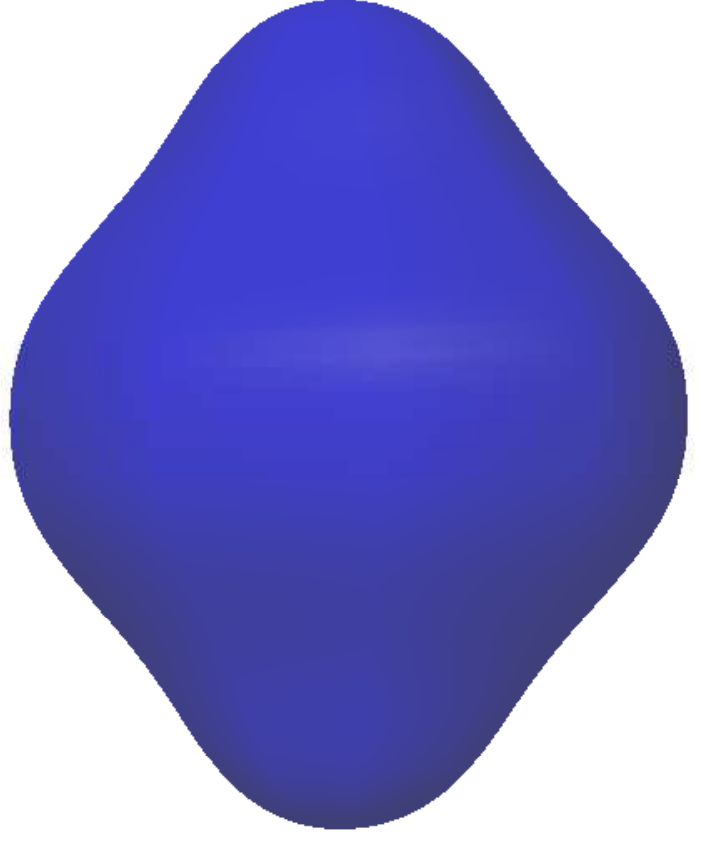}}
     \subfigure[t=1.26]{\includegraphics[viewport=-25 -75 275 225,angle=0,scale=0.42]{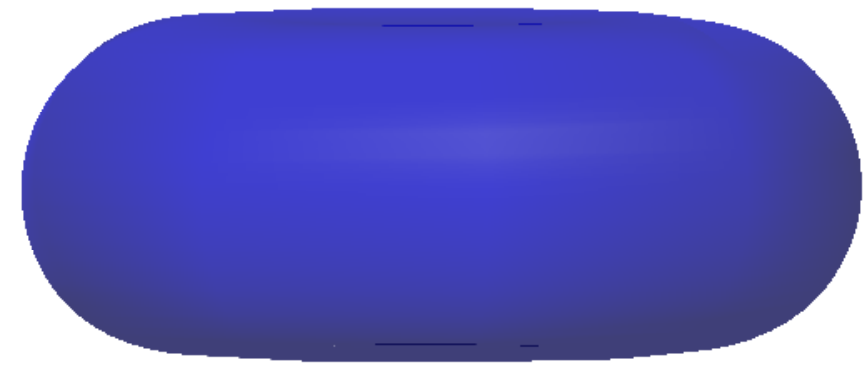}}
    \end{minipage}
    \begin{minipage}[c]{.45\textwidth}     \centering
     \subfigure[t=1.68]{\includegraphics[viewport=-25 -75 275 225,angle=0,scale=0.42]{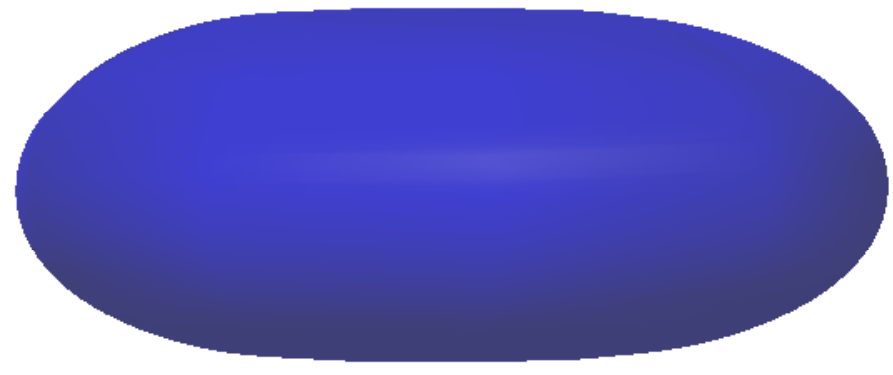}}
     \subfigure[t=2.09]{\includegraphics[viewport=-60 -20 240 280,angle=0,scale=0.42]{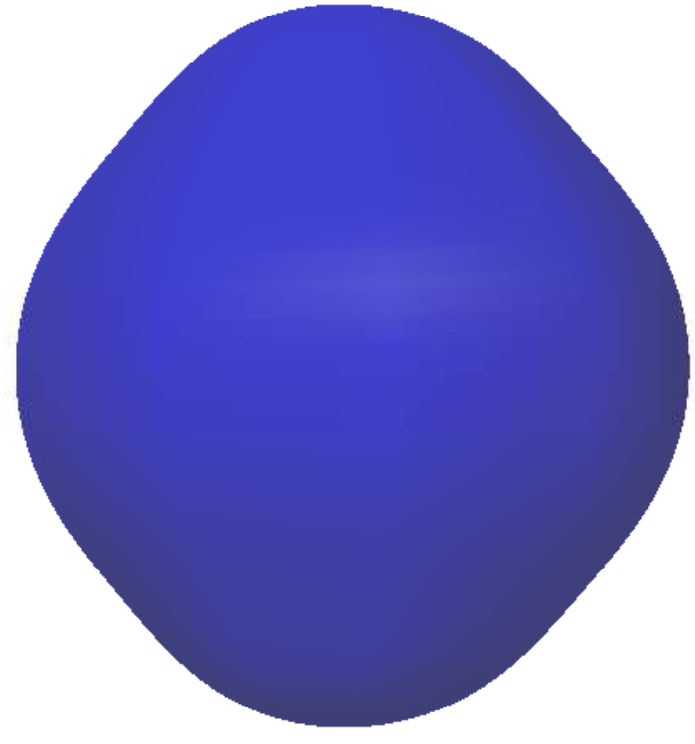}}
     \subfigure[t=2.51]{\includegraphics[viewport=-75  0  225 300,angle=0,scale=0.42]{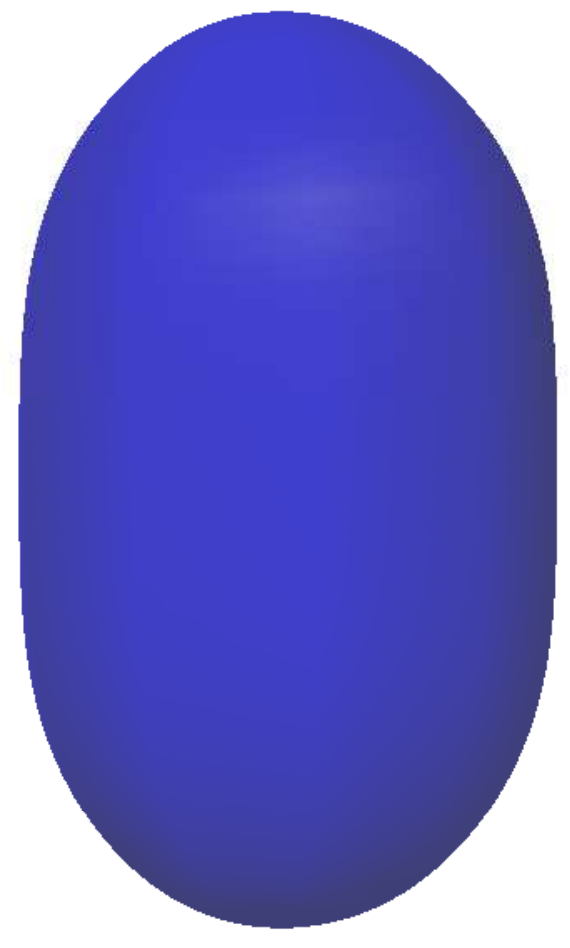}}\\
     \subfigure[t=2.93]{\includegraphics[scale=0.2]{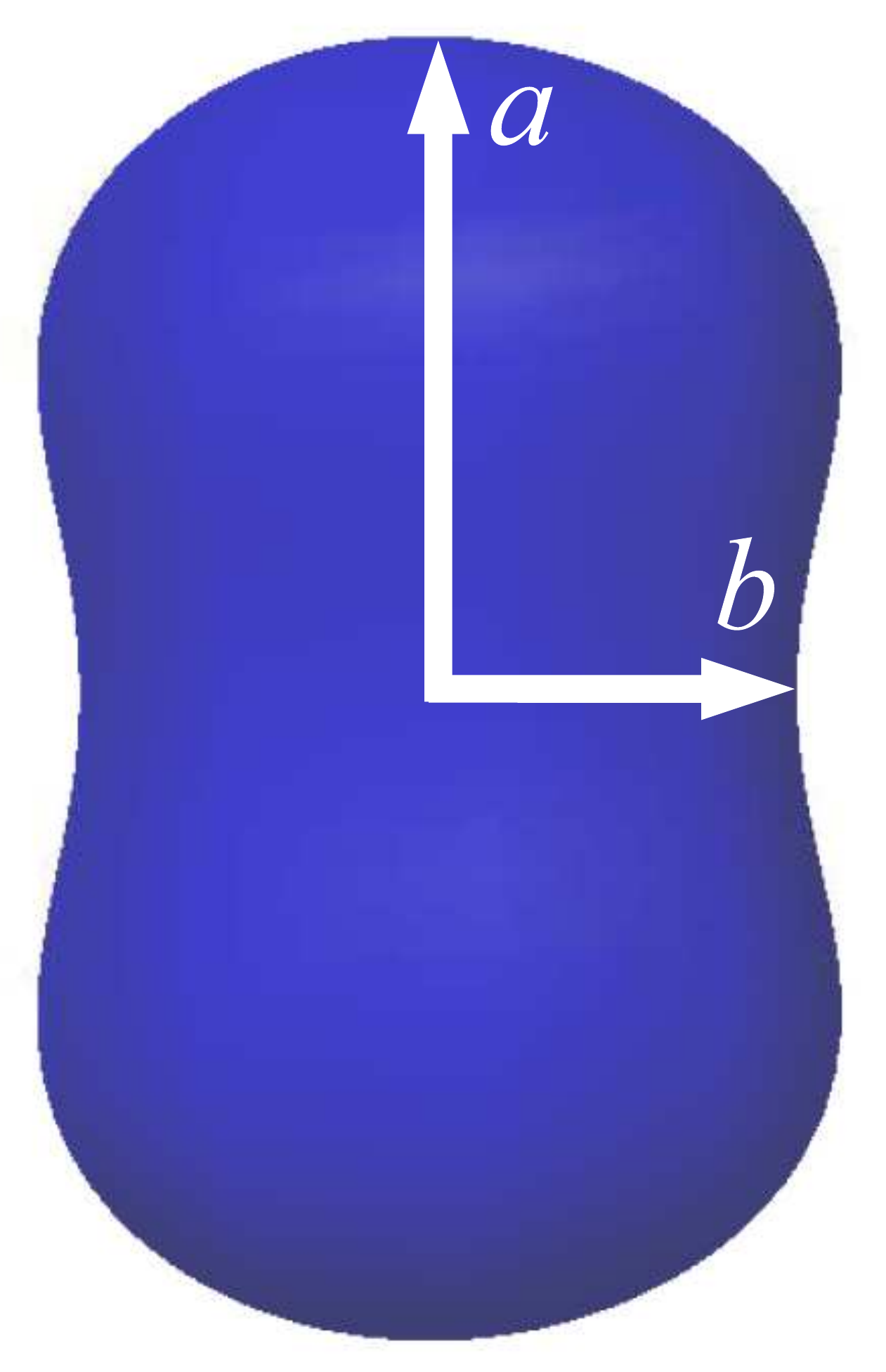}}
     \end{minipage}
\caption{Evolution of a liquid drop with Re=100 and We=1 over one period ($T=2.93$).}\label{F:oscdrop_f2_evo}
\end{figure}

Parameters are chosen to allow a comparison of our results with the numerical studies reported by other groups in \citep{basaran92} and \citep{meradji01}. To do so, we run two simulations of the axisymmetric oscillation of a drop with $Re=10,100$  and $Ca=0.1,0.01$ (so that $We=Re\,Ca=1$ for both simulations), respectively.  We consider the oscillation of a viscous liquid droplet whose initial shape is most naturally represented in spherical polar coordinates $(R,\alpha,\varphi)$, with the
origin located at the centre of the drop so that the drop surface $\mathbf{x}_G$ is given by
\begin{equation}\label{osc_shape} \mathbf{x}_G = f(\alpha,t) \mathbf{e}_{R},
\end{equation}
where $\mathbf{e}_{R}$ is a unit vector in the radial direction.  Then, as shown  in Figure~\ref{F:oscdrop_f2_evo}h, $a=f(0,t)$ is the length of the semi-major axis and $b=f(\pi/2,t)$ is the length of the semi-minor axis.

In the benchmark test case, the drop is released from a shape whose deviation from a sphere is proportional to the $2$nd spherical harmonic $P_2(\cos\alpha) = \hbox{$\frac{1}{2}$}(3\cos^2\alpha-1)$, with coefficient of proportionality chosen to be $0.9$, so that

\begin{equation}\label{osc_harmonic} f(\alpha,0) = \gamma [1+0.9~P_{2}(\cos\alpha)],
\end{equation}
where $\gamma$ is a normalizing factor which ensures that the droplet has the correct non-dimensional volume, in our case $\gamma=4\pi/3$.

We record the time $T_1$, and aspect ratio of the drop after one period $(a/b)_{T_1}$, for $Re=10,100$ and $We=1$ in order to validate our code against the previous studies. For the results which are presented, a relatively crude mesh of $630$ elements was used with a fixed time step of
$\delta t = 0.001$. Doubling the number of elements or reducing the time step by a factor of ten
resulted in a change of less than $0.1\%$ in both the time and amplitude recorded after one period. Significantly, the results in \onlinecite{basaran92} were for
$128$ elements whilst \onlinecite{meradji01} use an order of magnitude more elements.

\begin{figure}\centering
\includegraphics[scale=0.35]{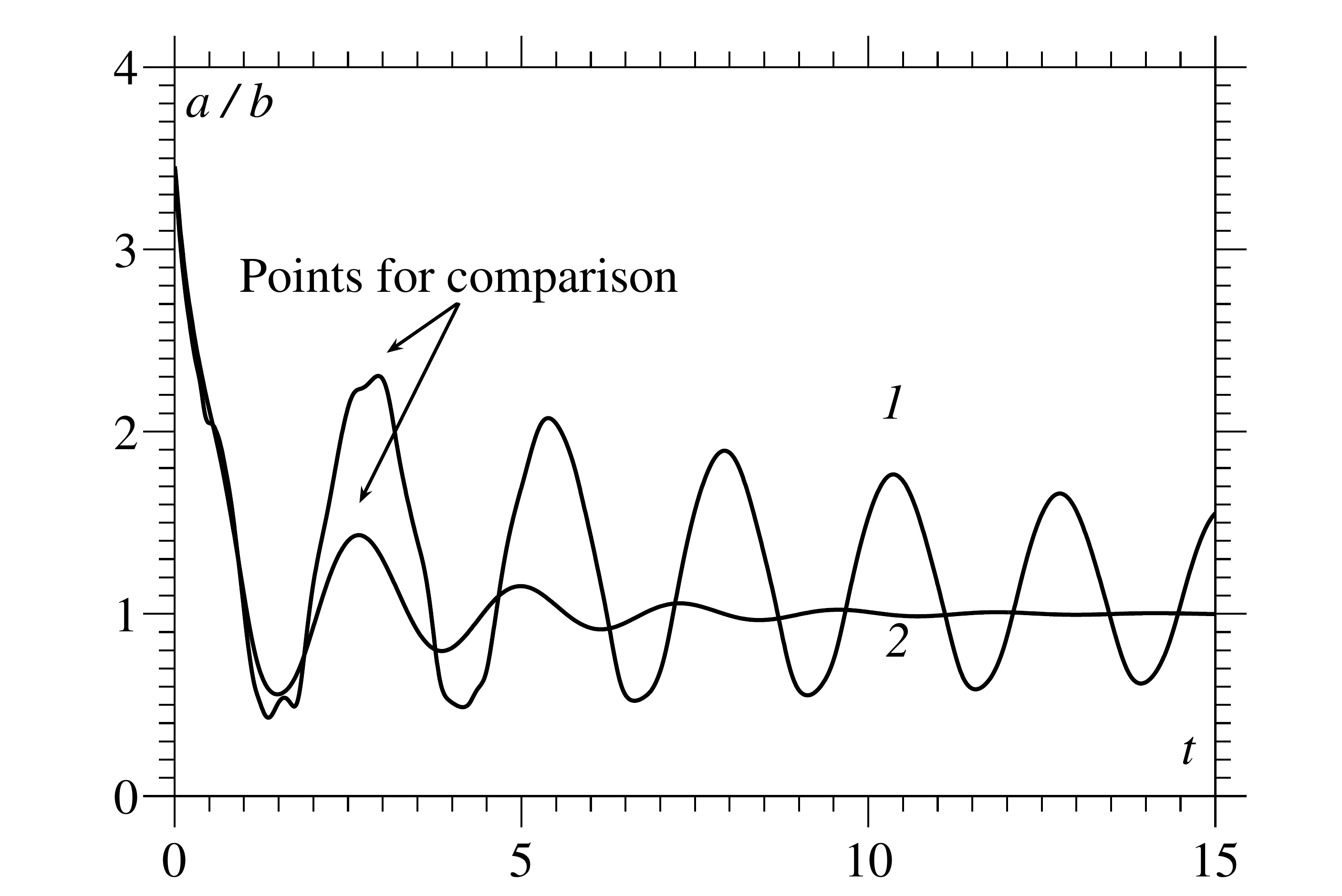}
\caption{Aspect ratio $a/b$ of two drops over a number of periods with
curve $1$ obtained using Re=100, We=1 whilst curve $2$ is for Re=10,
We=1.}\label{F:oscdrop_f2_aspect}
\end{figure}

Figure~\ref{F:oscdrop_f2_evo} shows the evolution of the drop over one period for the $Re=100$ case. The high deformation of the free surface is clear and one can observe that at the end of the first period the drop, whose equilibrium shape is a sphere, has its amplitude of oscillation damped by viscosity. Figure~\ref{F:oscdrop_f2_aspect} shows the time-dependence of the aspect ratio $a/b$, which is recorded after one period $t=T$. It should be pointed out that the kinks in curve 1 of
Figure~\ref{F:oscdrop_f2_aspect} are not numerical artifacts; they are associated with the high
deformation regime and can be seen in the previous studies.

Our results in Table~\ref{table} are seen to be in good agreement with both studies. The values align most
closely with those of \onlinecite{meradji01}, which is reassuring given the greater mesh resolution
associated with this study.
\begin{table}
\centering
\begin{tabular}{|c|c|c|c|}
\hline
         Re=10    & \onlinecite{basaran92} & \onlinecite{meradji01}  & Present Work \\
\hline
  $T_{1}$         & 2.660  & 2.640    & 2.656   \\
  $(a/b)_{T_{1}}$ & 1.434  & 1.432    & 1.432   \\
\hline
\hline
       Re=100     &         &         &        \\
\hline
 $ T_{1} $        & 2.905   & 2.930   & 2.936  \\
 $ (a/b)_{T_{1}}$ & 2.331   & 2.304   & 2.305  \\
\hline
\end{tabular}
\caption{Comparison of current results with previous studies.}\label{table}
\end{table}

Thus, it has been demonstrated that our numerical framework is able to provide accurate results for complex unsteady free-surface flows. The oscillation of liquid drops is a problem of interest in its own right, and at this point we could look at comparing our results to experiments in the literature \citep{wang96}, to probe newly proposed analytic models for decay rates \citep{smith10} or to consider the influence that including interface formation physics may have when oscillations are of high frequency and the interface is forming and disappearing at a significant rate. All of these avenues of investigation are being pursued but lie beyond the scope of the present paper, and we now turn our attention to the code's capabilities at describing drop impact and spreading phenomena.

\section{Microdrop Impact and Spreading}\label{drops}

To study the key physical effects in microdrop impact and spreading phenomena in a range applicable to inkjet printing technologies, we consider simulations of impact on hydrophilic, hydrophobic and patterned substrates. In particular, we see that our code is able to account for the two extreme outcomes of the drop impact and spreading process, i.e. deposition and rebound; to recover information about the drop's dynamics which is currently unobtainable from purely experimental analysis; and to predict new methods for flow control on chemically patterned surfaces.

Consider a microdrop of water of radius $L=25$~$\mu$m which impacts a solid substrate at $U=5$~m~s${}^{-1}$, with the subsequent motion considered axisymmetric. Then, the non-dimensional parameters are $Re=130$, $Ca= 0.07$, $St=0.001$ and estimates for the interface formation model's parameters are taken from \onlinecite{blake02}. All that remains to be specified is the wettability of the solid substrate, which is characterized by the equilibrium contact angle $\theta_e$ that a free surface forms with the solid.

\begin{figure}
     \centering
     \vspace{-2.7cm}
     \begin{minipage}[l]{.49\textwidth}
     \vspace{-0.2cm}
\subfigure[t=0]{\includegraphics[scale=0.35]{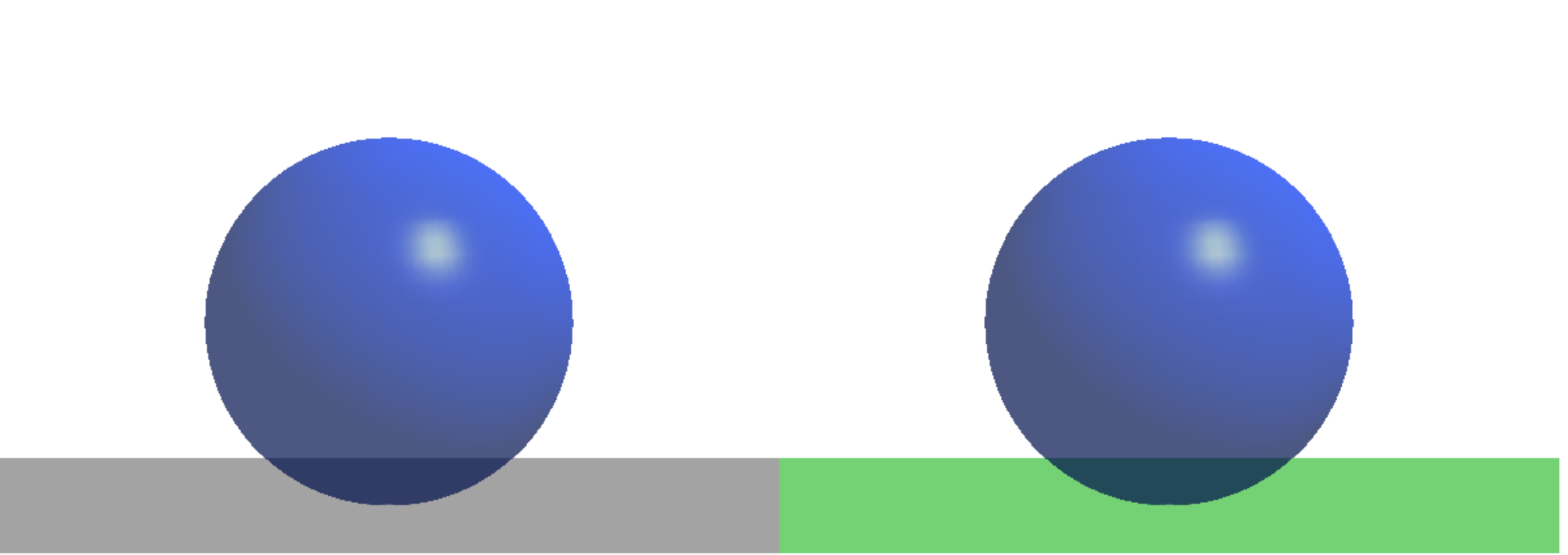}}
\subfigure[t=0.2]{\includegraphics[scale=0.35]{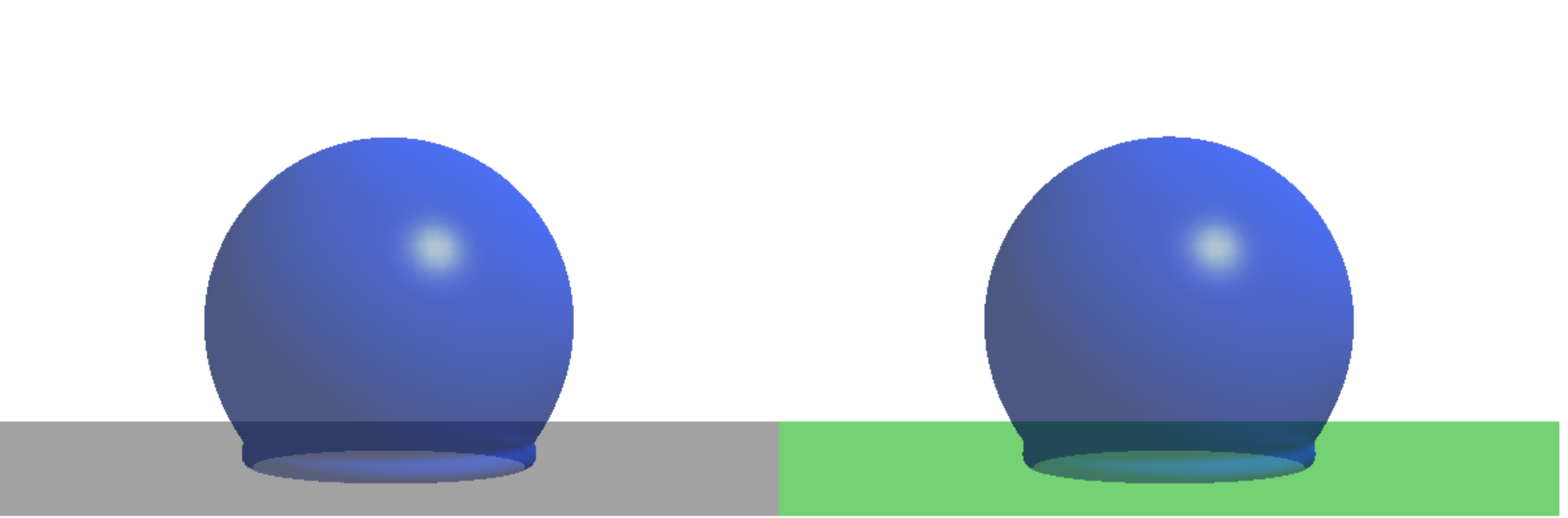}}
\subfigure[t=0.4]{\includegraphics[scale=0.35]{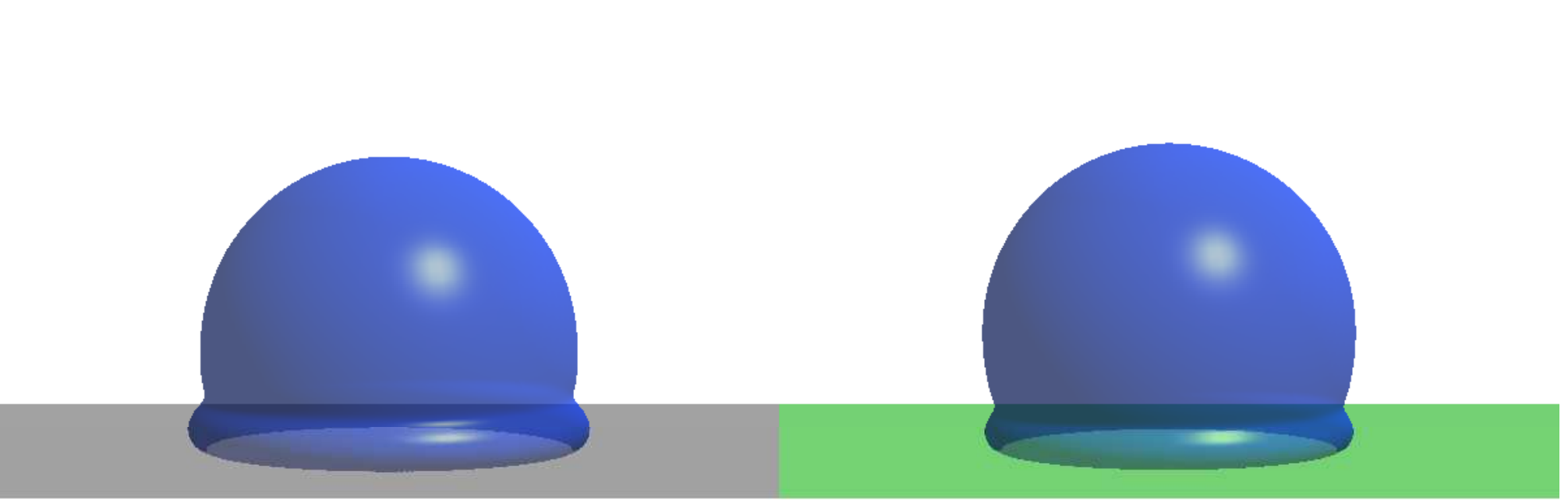}}
\vspace{0.6cm}
\subfigure[t=0.6]{\includegraphics[scale=0.35]{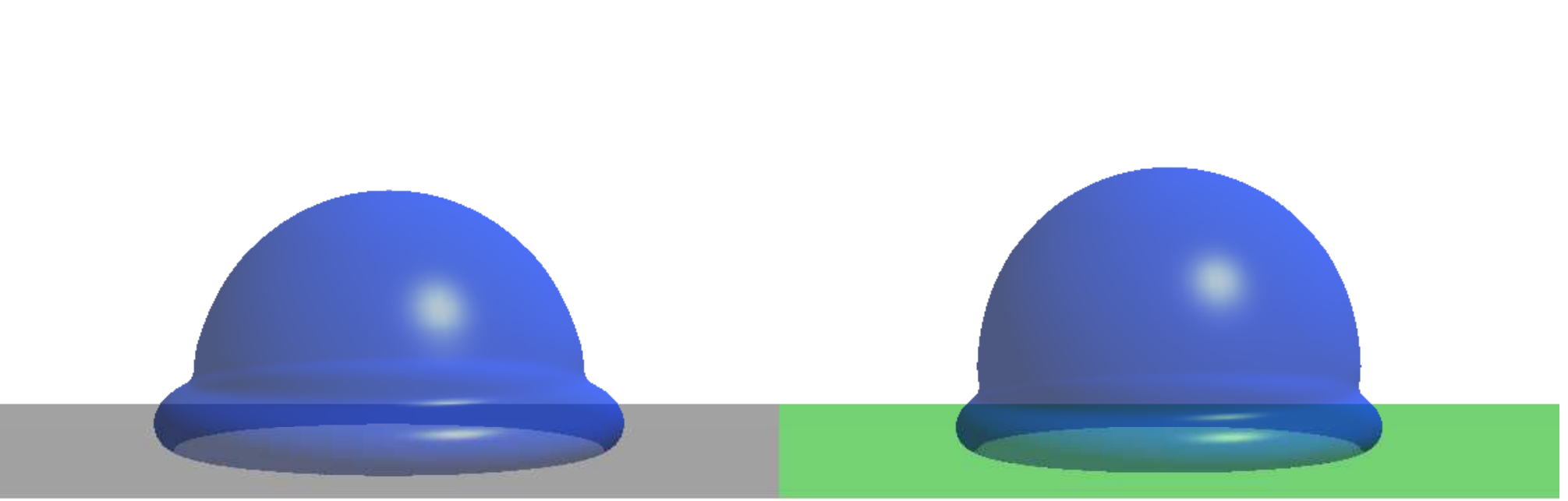}}
\vspace{1cm}
\subfigure[t=0.8]{\includegraphics[scale=0.35]{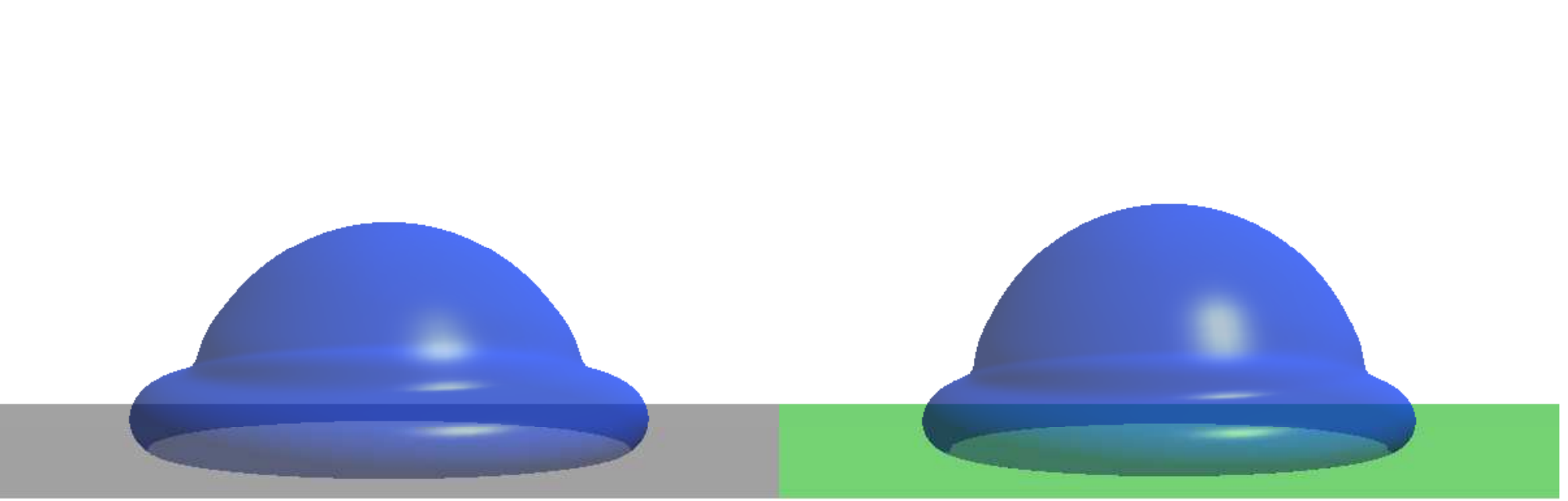}}
\subfigure[t=1.2]{\includegraphics[scale=0.35]{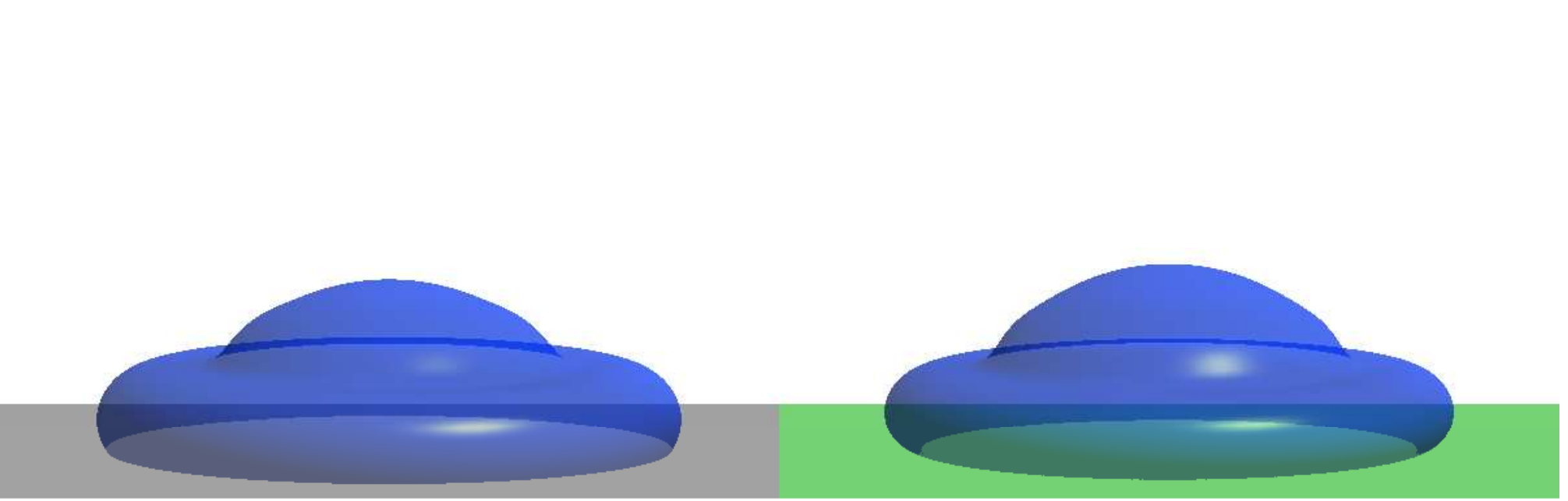}}
    \end{minipage}
    \begin{minipage}[r]{.49\textwidth}
\subfigure[t=1.6]{\includegraphics[scale=0.35]{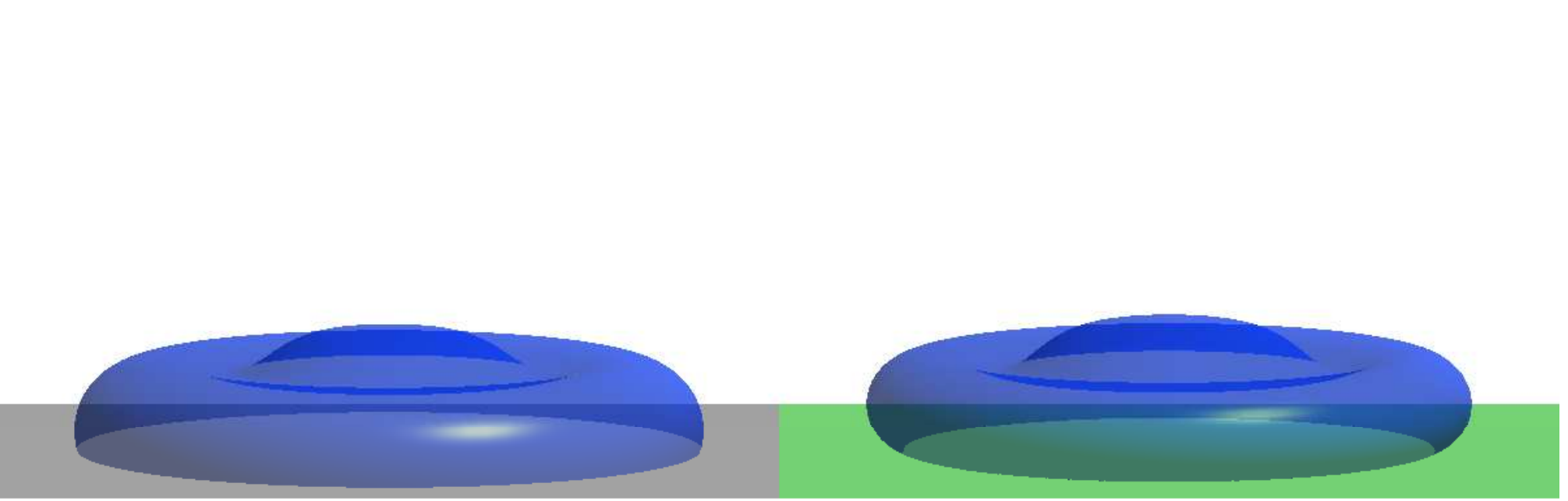}}
\subfigure[t=2]{\includegraphics[scale=0.35]{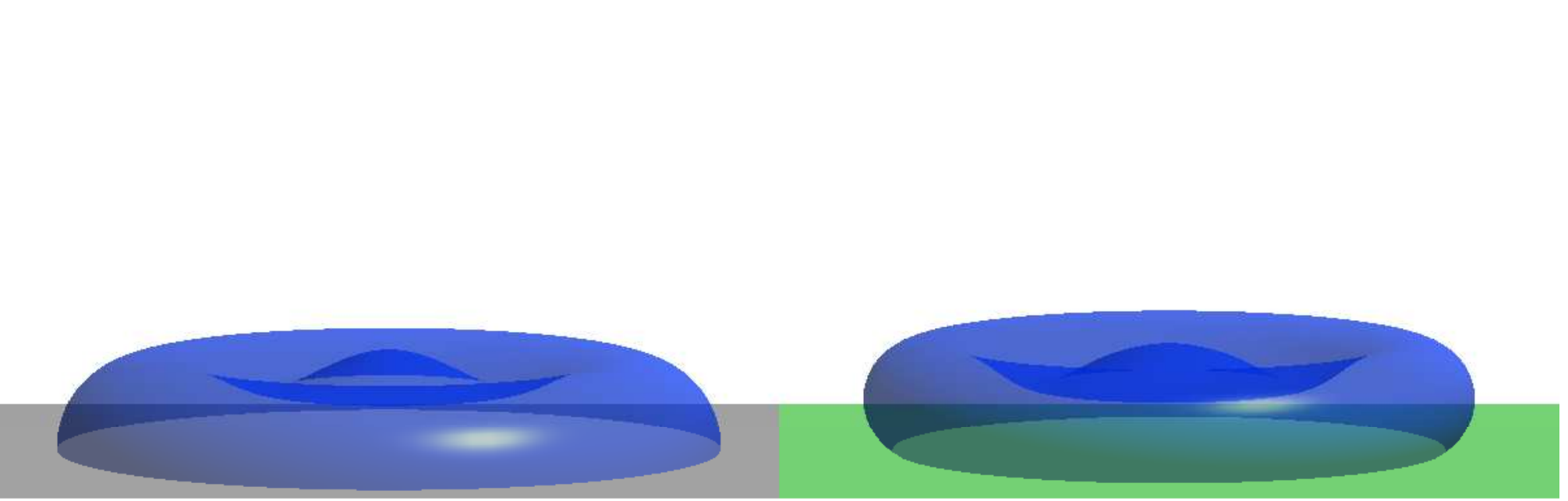}}
\subfigure[t=3]{\includegraphics[scale=0.35]{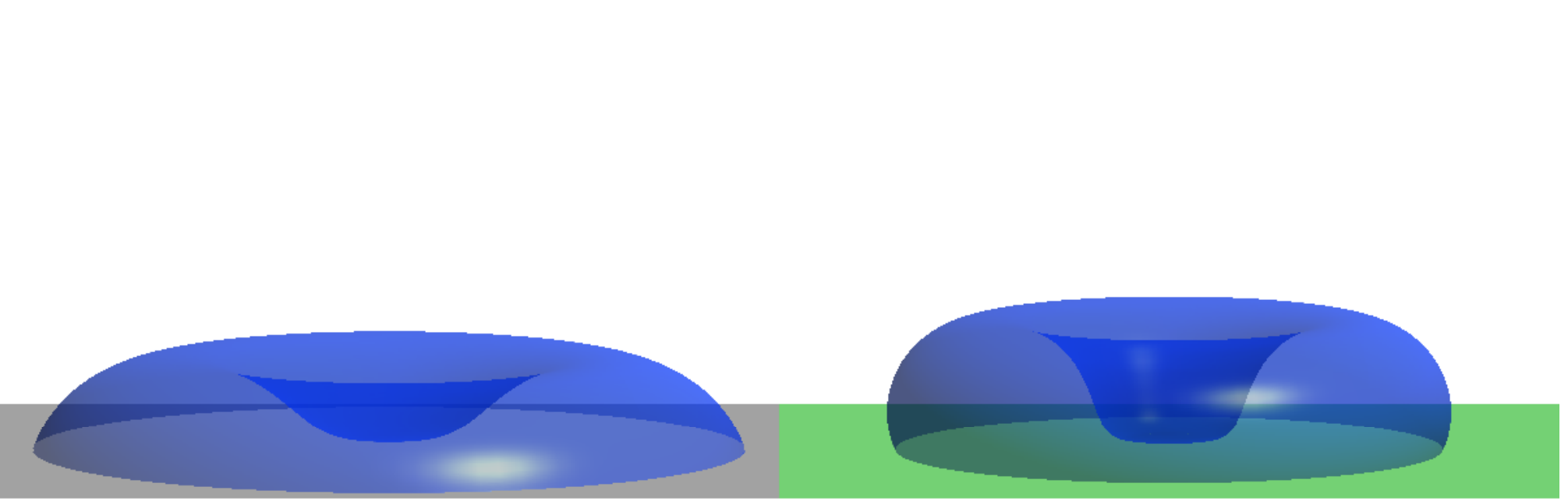}}
\subfigure[t=4]{\includegraphics[scale=0.35]{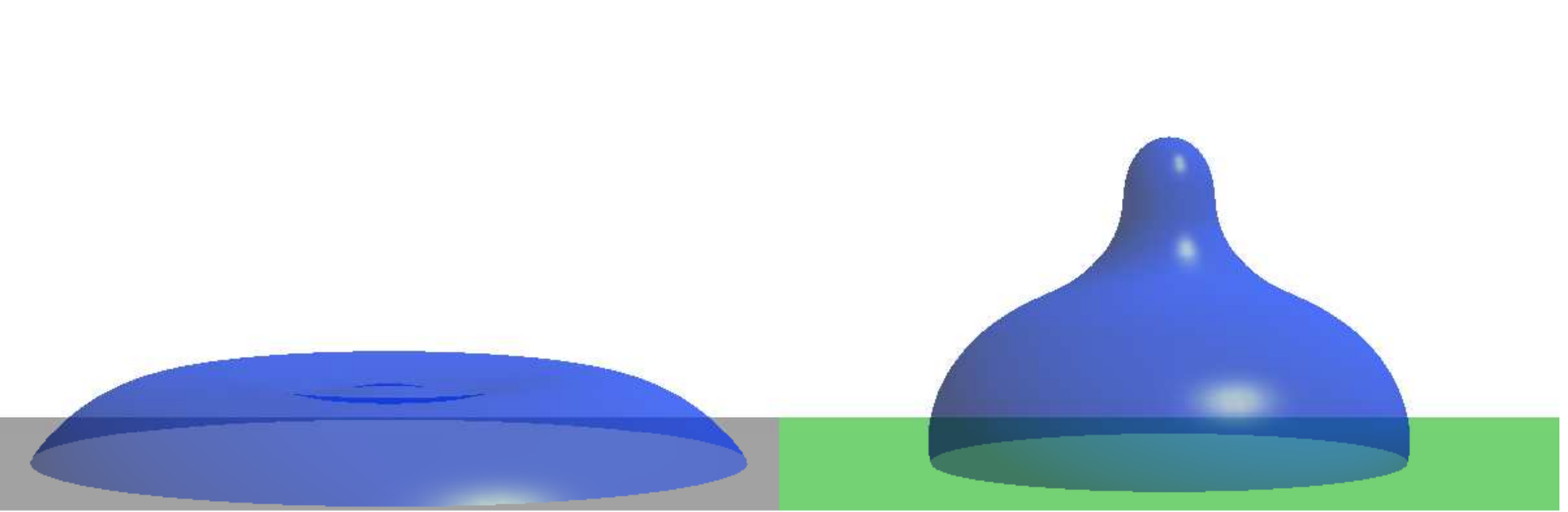}}
\subfigure[t=6]{\includegraphics[scale=0.35]{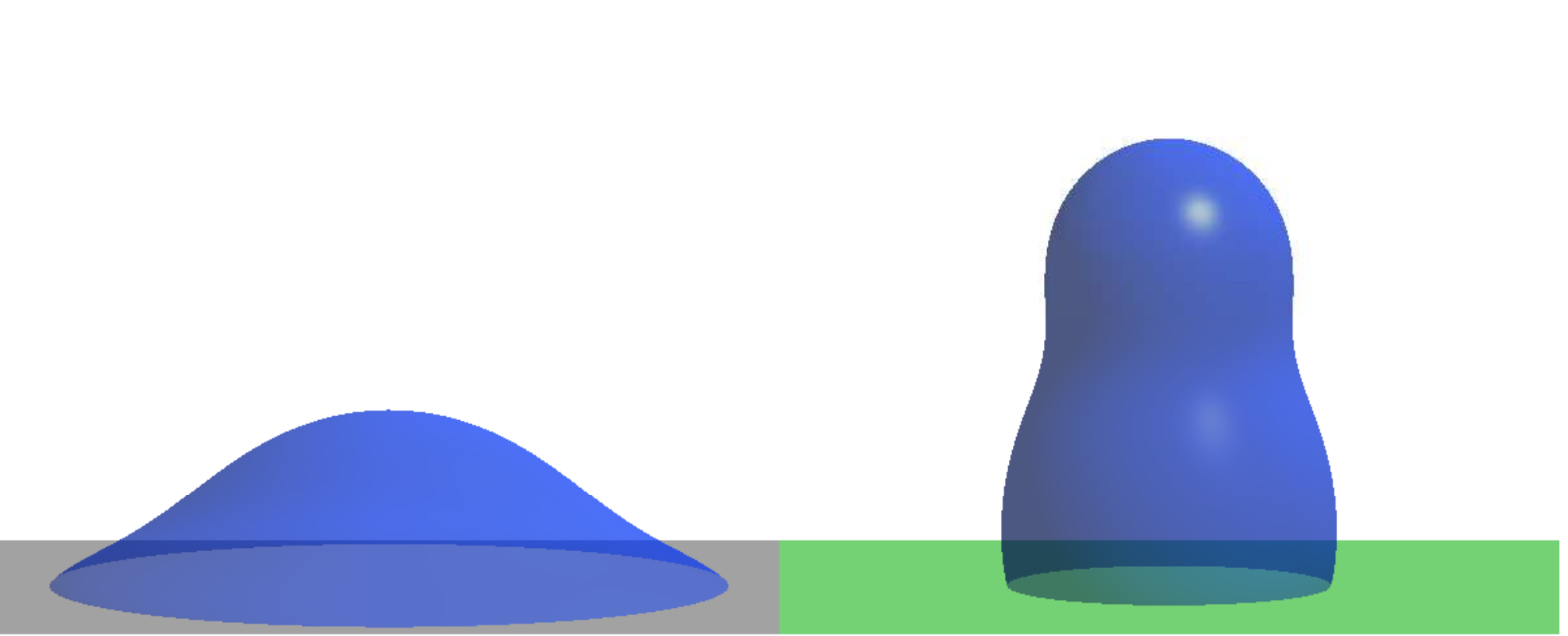}}
\subfigure[t=10]{\includegraphics[scale=0.35]{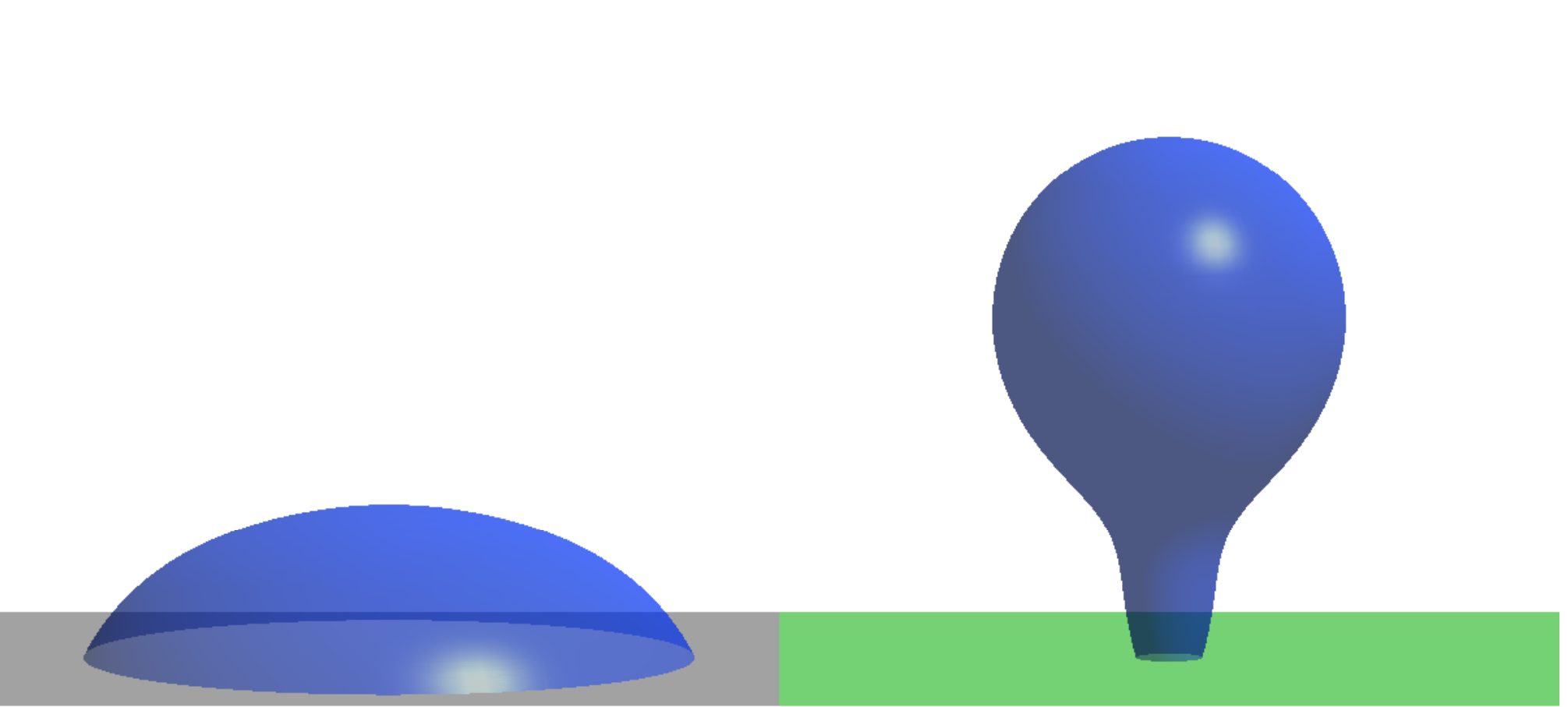}}
     \end{minipage}
\caption{Microdrop impact and spreading simulations at $Re=130$, $Ca= 0.07$, $St=0.001$. The grey substrate (left) is hydrophilic ($\theta_e=60^\circ$) whilst the green one (right) is hydrophobic ($\theta_e=130^\circ$).}\label{F:drops}
\end{figure}

\begin{figure}\centering
\includegraphics[scale=0.32]{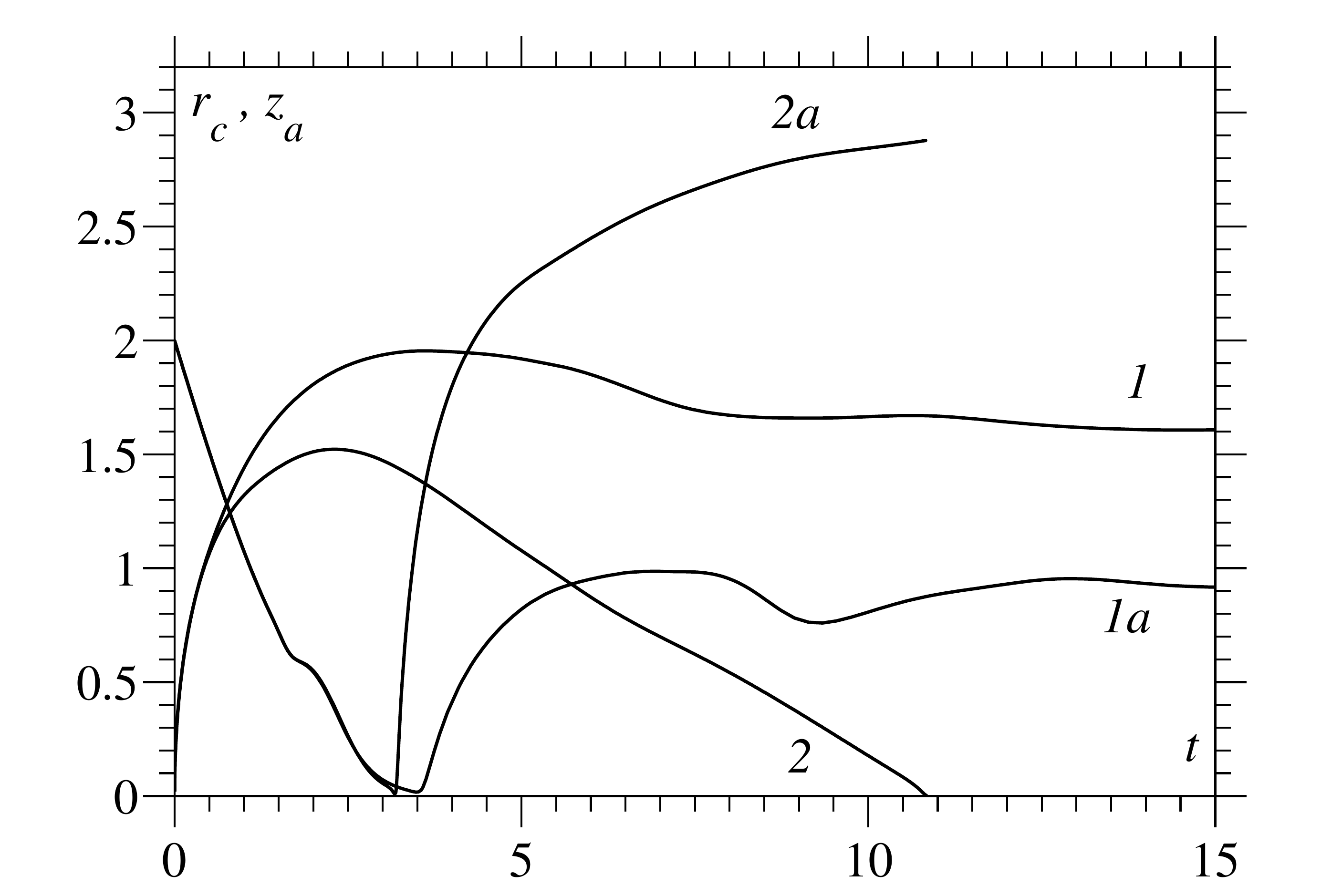}
\caption{Evolution of the drop's contact line position $r_c$ ($1,~2$) and apex height $z_a$ ($1a,~2a$) as a function of time for simulations at $Re=130$, $Ca= 0.07$, $St=0.001$. Curves $1$ \& $1a$: $\theta_e=60^\circ$ and curves $2$ \& $2a$: $\theta_e=130^\circ$.}\label{F:drops_position}
\end{figure}

Two simulations are shown in Figure~\ref{F:drops} for the impact of a drop on substrates of different wettabilities,  $\theta_e=60^\circ$ and $\theta_e=130^\circ$, respectively, with an associated link to a movie here: \href{http://www.youtube.com/watch?v=uFP-4EY8Ekw&feature=youtu.be}{Movie}. The evolutions of the contact line radius $r_c$ and and apex height $z_a$ are given in Figure~\ref{F:drops_position}.   Hereafter, for brevity we use the term `apex' for the point located at the intersection of the axis of symmetry and the free surface; as Figure~\ref{F:apex} shows, it isn't necessarily the highest point of the free surface.

\begin{figure}
     \centering
\subfigure[t=0.5]{\includegraphics[scale=0.35]{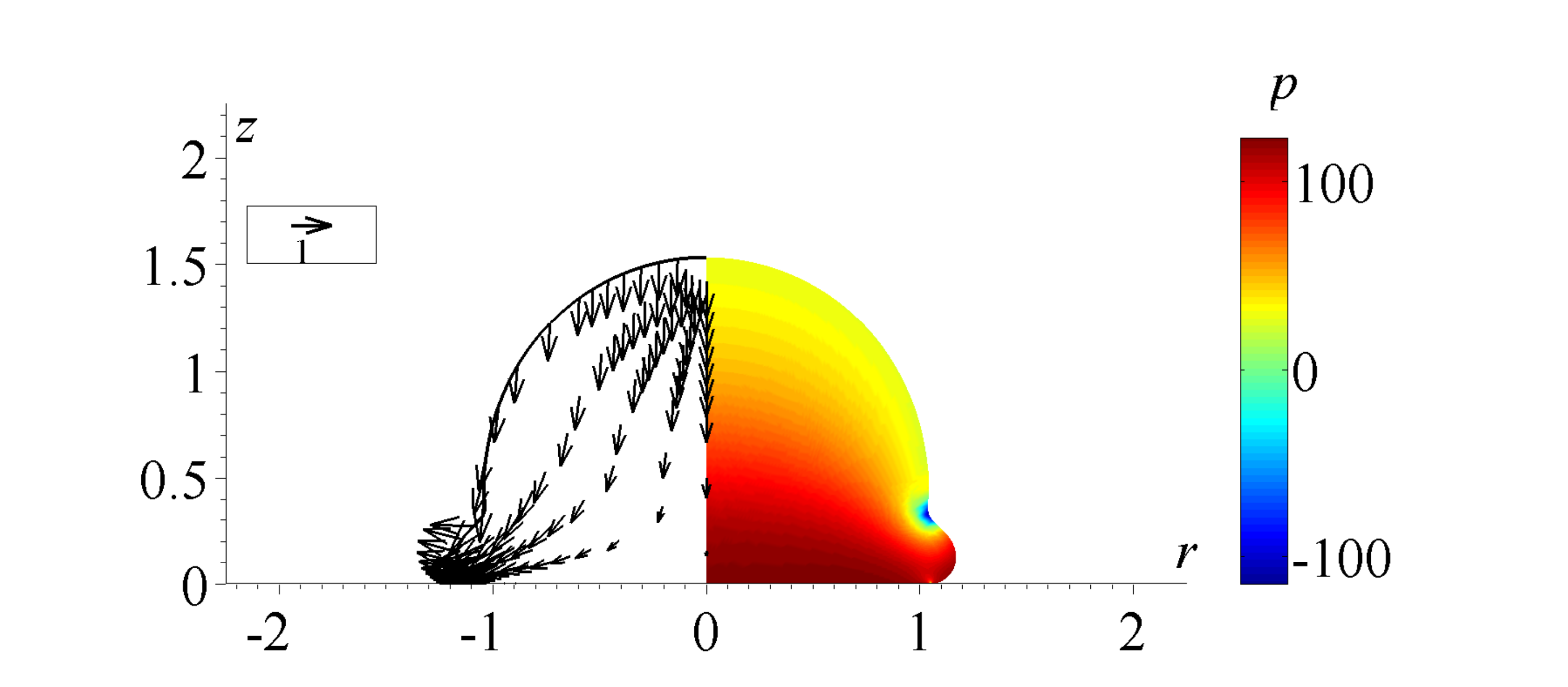}}
\subfigure[t=1]{\includegraphics[scale=0.35]{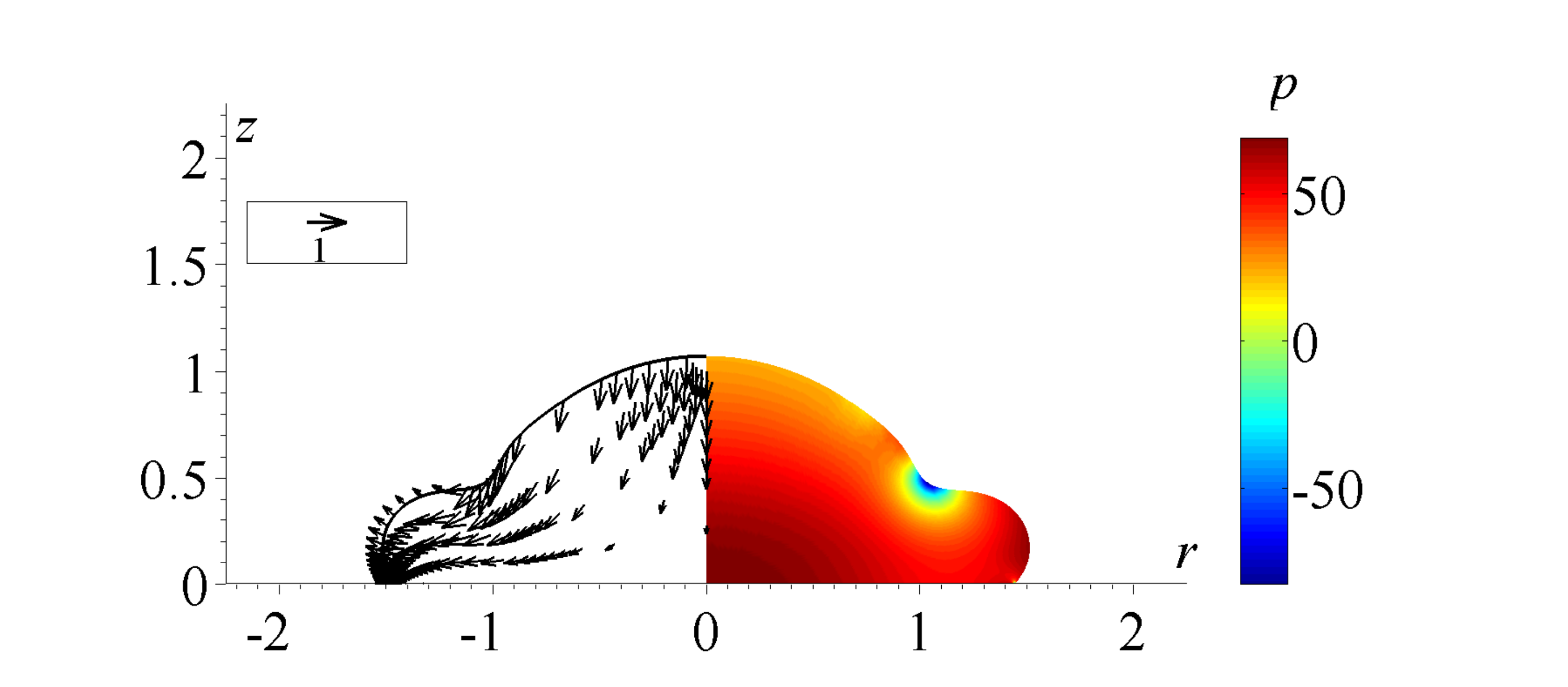}}
\subfigure[t=3]{\includegraphics[scale=0.35]{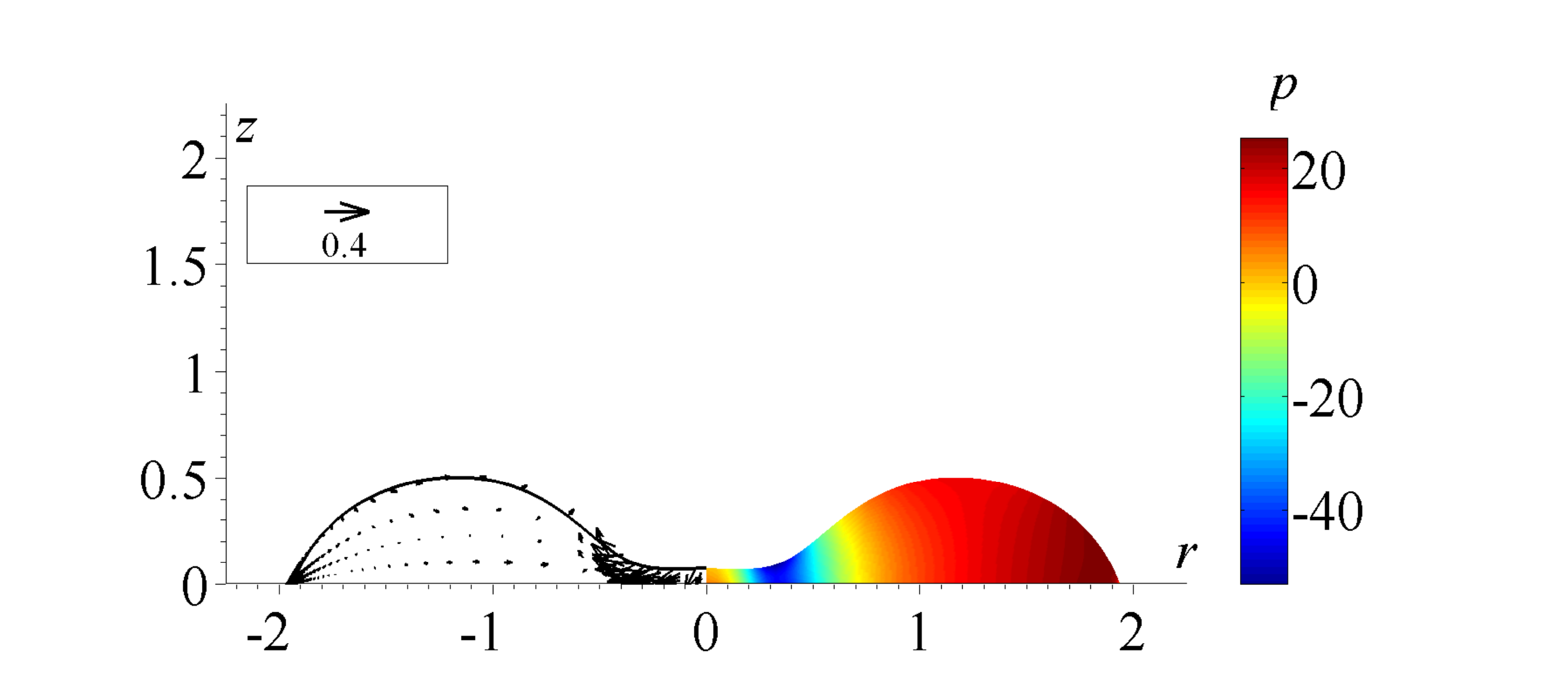}}
\caption{Microdrop impact and spreading simulation at $Re=130$, $Ca= 0.07$, $St=0.001$ on a hydrophilic substrate  ($\theta_e=60^\circ$). Left: Velocity vectors, Right: Pressure field.}\label{F:drops_colour}
\end{figure}

In Figure~\ref{F:drops} it can be seen that in the early stages of spreading when inertia is dominant, roughly until $t=1$ (with one dimensionless unit corresponding to $5$~$\mu$s for the drop on which our non-dimensional parameters were based), the shapes of the two drops are indistinguishable. As can be seen in Figure~\ref{F:drops_colour}, the contact line is forced outwards as fluid is pushed out radially from the centre of the drop which is being driven vertically into the substrate by inertial forces. This causes a toroidal rim of fluid to form near the contact line, with the pressure plot in Figure~\ref{F:drops_colour} clearly showing the formation of a disturbance, travelling from the contact line towards the apex, which separates the growing rim of fluid from the bulk of the drop. Eventually, the drop starts to feel the wettability of the solid on which it is spreading and, since in both cases inertia has carried the drop past its equilibrium position, the contact line starts to recede. Noticeably, as can be seen in both Figure~\ref{F:drops} and Figure~\ref{F:drops_position}, although the wettability of the substrate eventually begins to alter the position of the contact line; this is not felt by the apex until a much later time. In fact, the initial fall of the apexes are very similar, and it is only upon their recoil, around $t=4$, that their paths begins to differ. When the drops begin to recoil, their motions differ quite significantly; notably, for the drop on the hydrophobic substrate, where the dewetting of the substrate occurs so quickly that the drop rebounds back off the substrate. By comparing the images at $t=3$ and $t=4$ in Figure~\ref{F:drops}, one can see that the rebound is preceded by a jet emanating from the apex region which pulls fluid radially inwards towards the axis of symmetry. This second stage of spreading is seen to be on a much longer time scale than the initial stages after impact.  The simulation has to be terminated as the drop is about to leave the substrate; extending the numerical platform to account for such behaviour is certainly viable. It is of interest to see that the drop's final shape is pear shaped and, indeed, this shape has been observed experimentally \citep{mao97}.

\begin{figure}\centering
\includegraphics[scale=0.32]{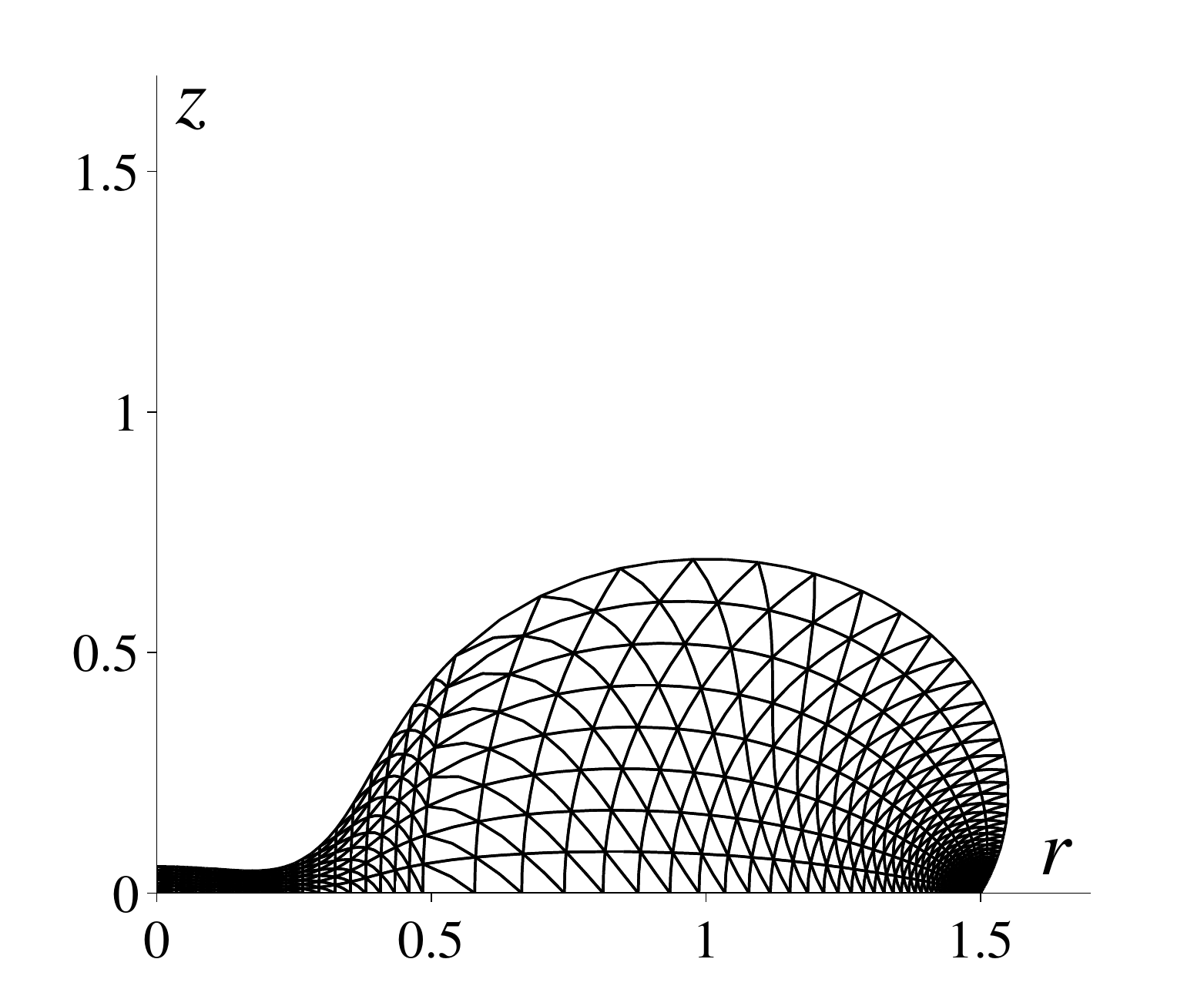}
\includegraphics[scale=0.32]{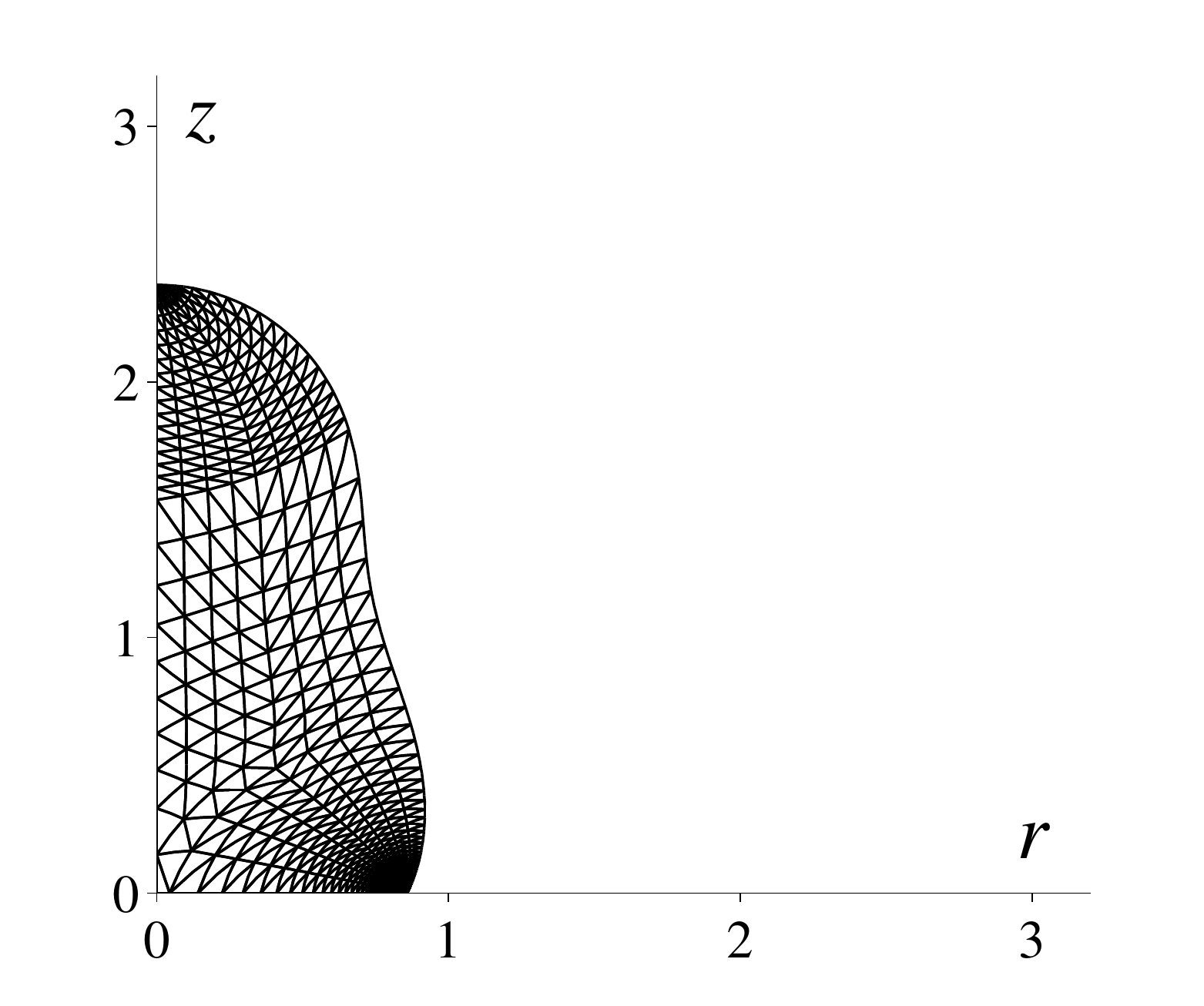}
\includegraphics[scale=0.32] {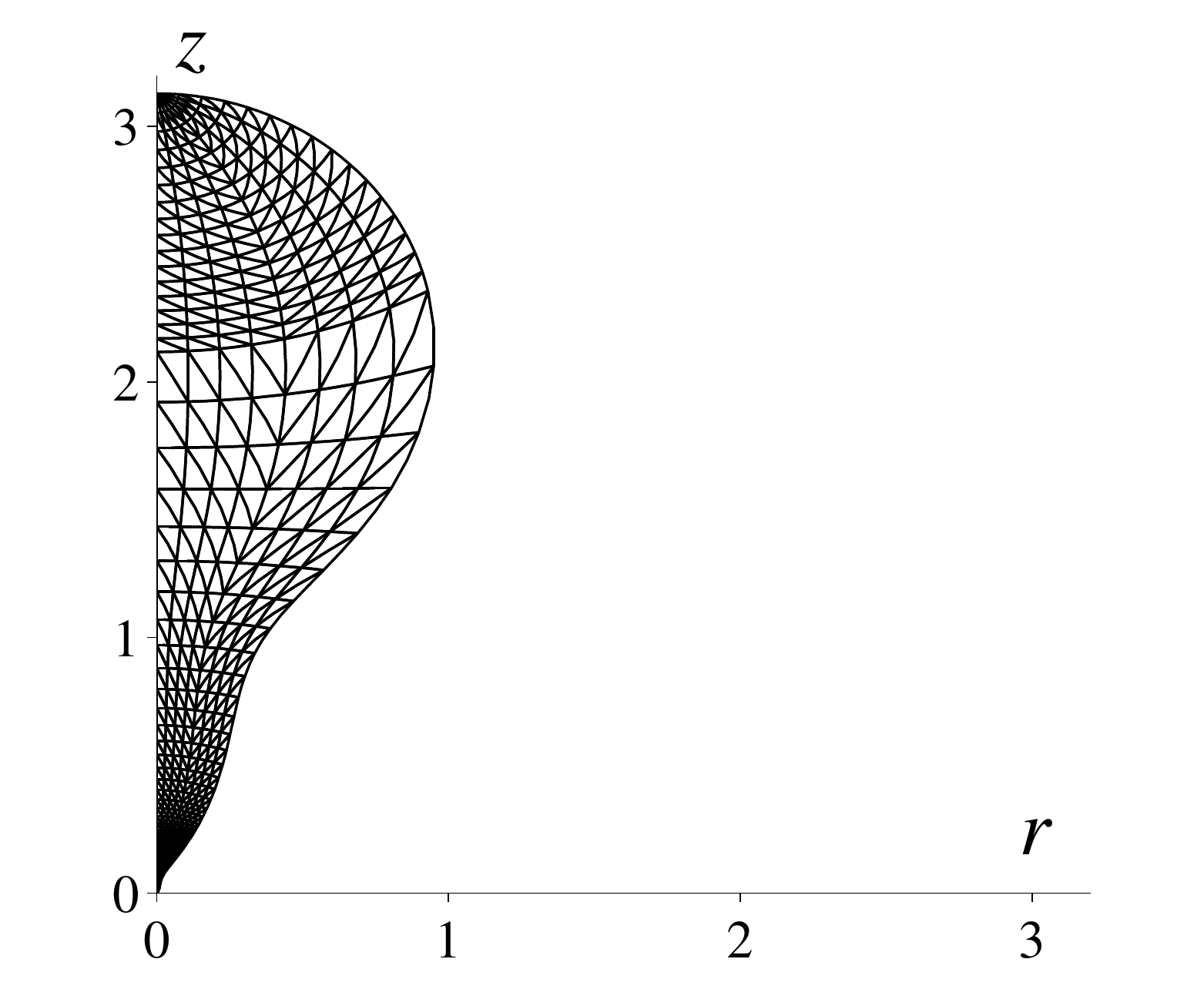}
\caption{Snapshots of a relatively coarse computational mesh during the impact, spreading and rebound
of the drop in Figure~\ref{F:drops} off a hydrophobic substrate at times $t=3,6,10.5$.  Enhanced online.}\label{F:rebound_mesh}
\end{figure}

\begin{figure}\centering
\includegraphics[scale=0.75]{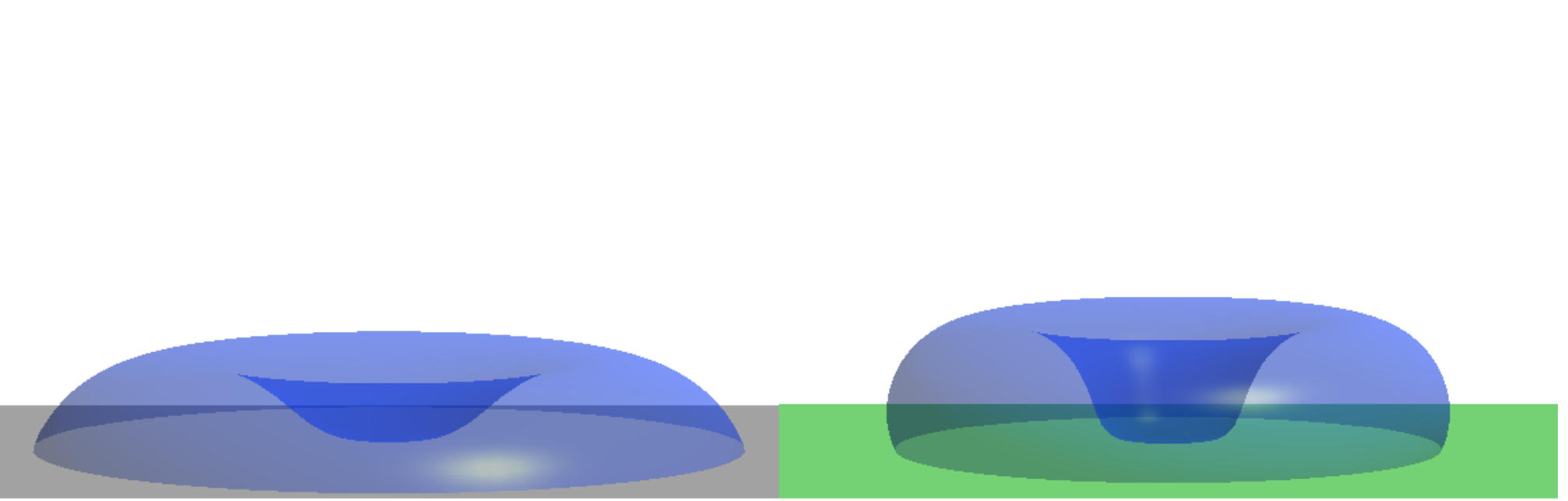}
\caption{Shape of the microdrops at $t=3$ highlighting dynamics which are inaccessible to experiments.}\label{F:apex}
\end{figure}

Snapshots of the mesh used during the computation of the obtained results are given in Figure~\ref{F:rebound_mesh}, where a relatively crude mesh is shown, allowing the elements to be easily identified. One can see that the mesh remains regular throughout extreme changes in free surface shape, like at $t=3$, when the apex is very close to touching the substrate, and at $t=10.5$, when the drop is close to leaving the substrate.

Some of our computational framework's advantages over a purely experimental analysis of the phenomenon are as follows. First, it can recover information which is inaccessible to experiments; second, one can efficiently map the influence of the system's parameters on the drop's dynamics, and, third, it is easy to attempt new things without the cost of full scale laboratory experiments.

As an illustration of the first of these advantages, as shown in Figure~\ref{F:drops_colour} and Figure~\ref{F:apex}, in our simulations we are able to see the entire shape of the microdrop for the whole simulation, whereas experimental images on microdrops are unable to show the dynamics of the apex as it disappears below the rim of fluid which surrounds it, as well as features experimentally unobtainable at these scales such as the flow field and pressure distribution inside the drop.  It can be seen that the apex gets extremely close to touching the solid substrate, i.e. to dewetting the centre of the drop.  This can also be seen from curve~$2a$ in Figure~\ref{F:drops_position}. The apex manages to recover just in time: as the contact line is receding, the apex re-emerges out of the centre of the drop in a jet-like protrusion (see Figure~\ref{F:drops}j).  One could envisage that if the drop's apex does dewet the centre of the drop, then this additional dissipation of energy may inhibit the rebound of the drop off the hydrophobic surface, although, initial indications in the relatively narrow parameter space investigated thus far show that microdrops are remarkably resilient to this dewetting feature. Understanding, and hence being able to control, this feature would be of significant benefit to processes which are constrained to using hydrophobic surfaces but would like to inhibit rebound. This aspect of the drop impact phenomenon will be the subject of future investigation.

With regard to the second advantage of reliable numerical simulations of microdrops over experiment, determining for what parameter values a drop will rebound is an important piece of information, particularly when the substrate to be used in a given process has to be hydrophobic, and it is difficult to find this out from experiments where one cannot vary material parameters of the system independently. With our numerical tool, parameter space can be mapped quickly and efficiently, so that one can ensure the drop deposits on a given substrate by, say, artificially changing the viscosity of the liquid or reducing the impact speed of the drop. Changing the substrate for the same microdrop impact parameters, we see that, for example, at $\theta_e = 110^\circ$ no rebound of the drop is observed.

Next, as an example of the code's cost-effectiveness with regard to the process, we consider the impact of drops on a chemically patterned surface which is custom built to enhance pre-determinable flow control on the drop's dynamics.

\section{Microdrop Impact and Spreading on Custom Built Chemically-Patterned Surfaces}\label{patt_surf}

\begin{figure}[h]
     \centering
     \begin{minipage}[l]{.49\textwidth}
\subfigure[t=0]{\includegraphics[scale=0.35]{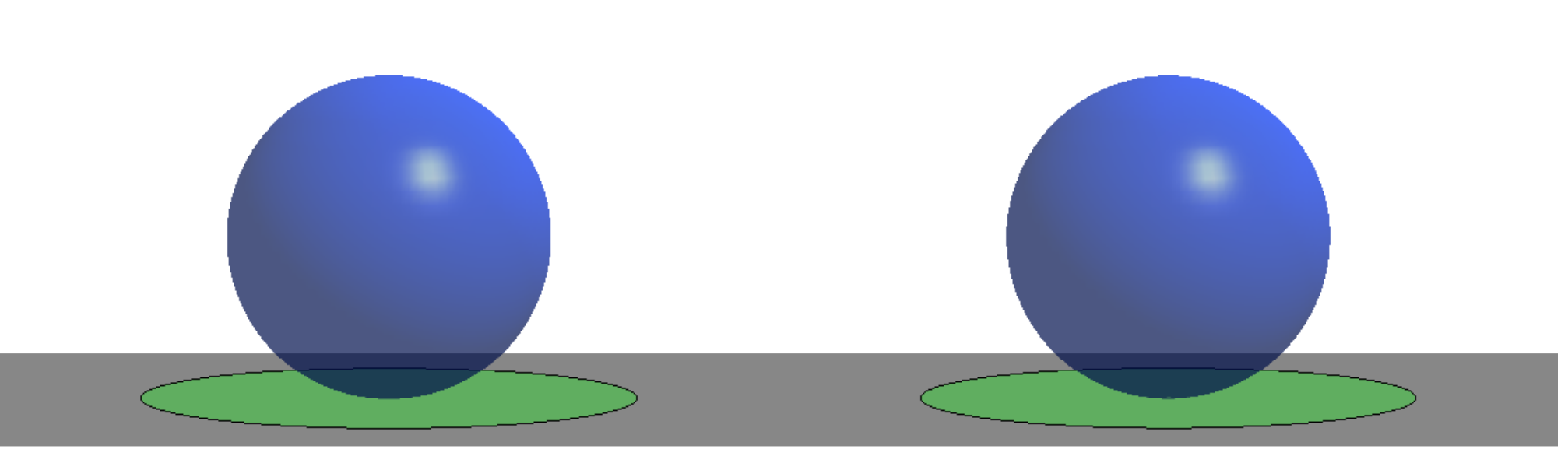}}
\subfigure[t=1]{\includegraphics[scale=0.35]{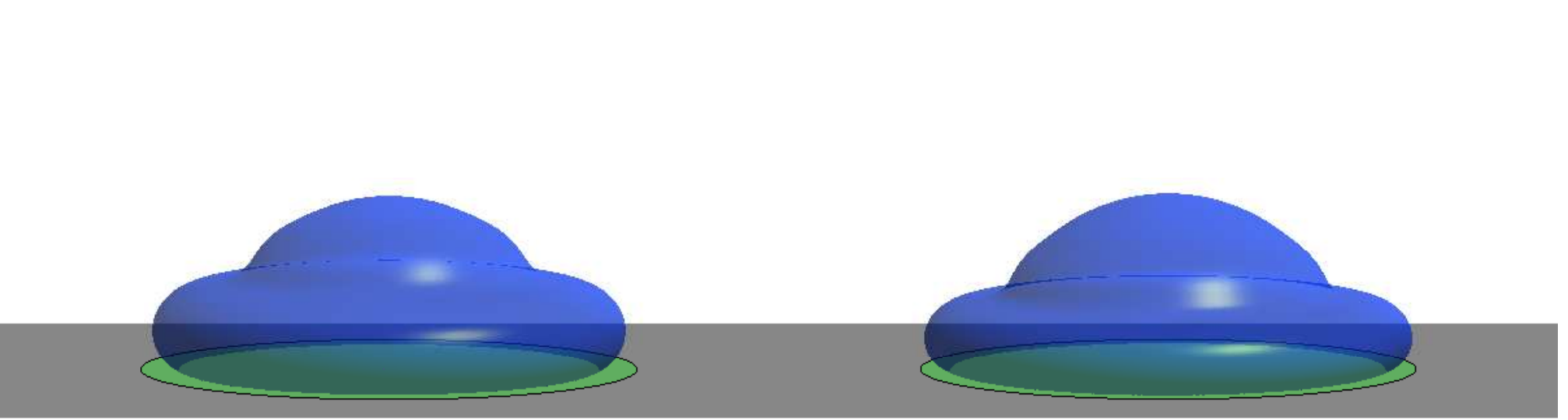}}
\subfigure[t=2]{\includegraphics[scale=0.35]{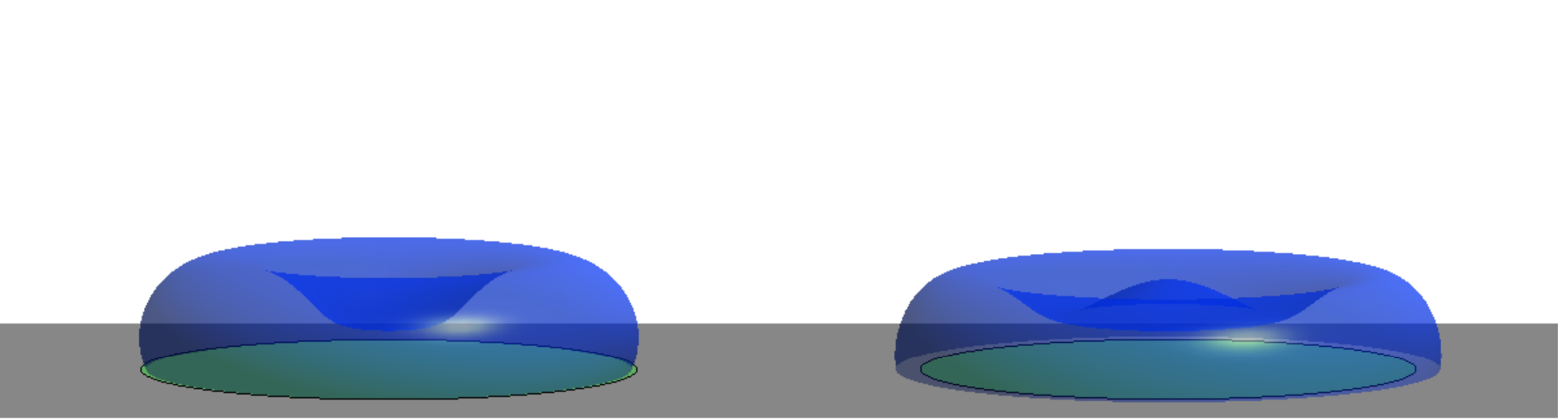}}
    \end{minipage}
    \begin{minipage}[r]{.49\textwidth}
\subfigure[t=3]{\includegraphics[scale=0.35]{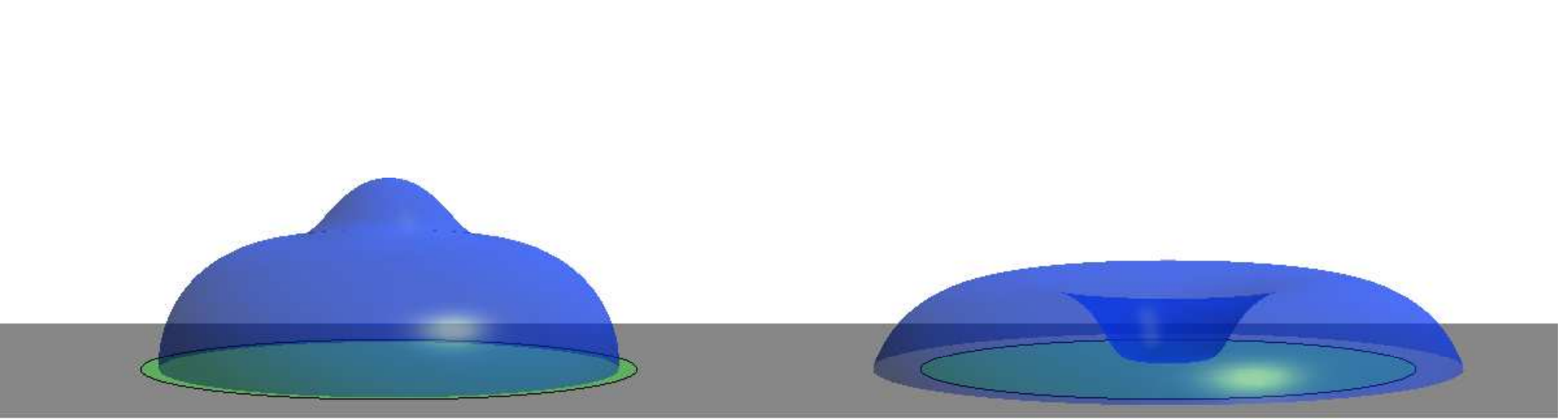}}
\subfigure[t=6]{\includegraphics[scale=0.35]{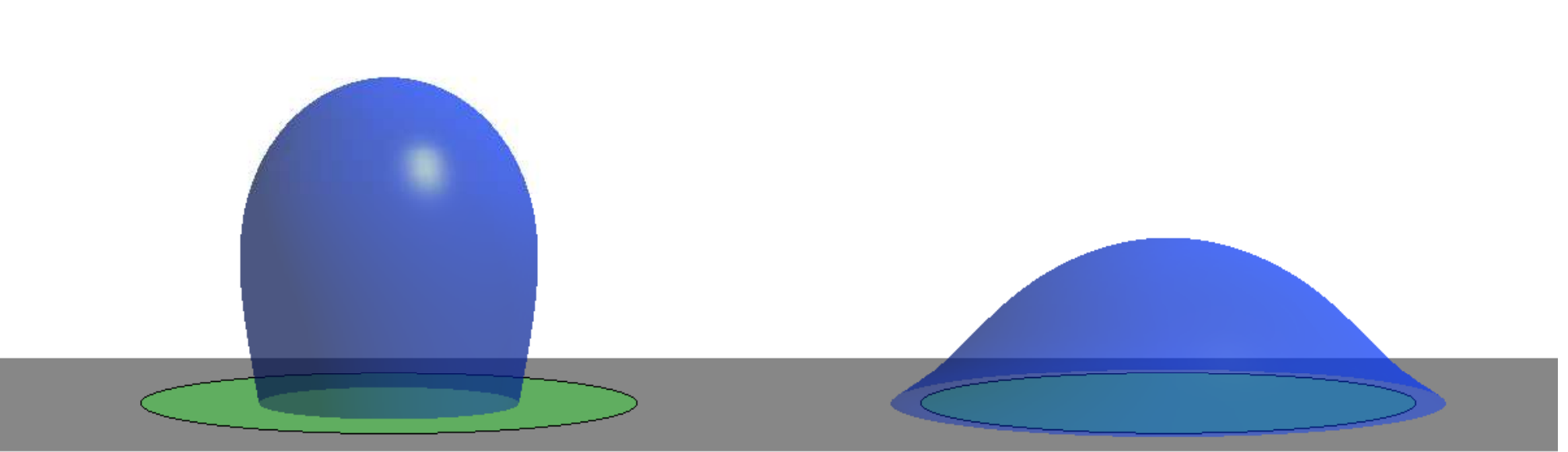}}
\subfigure[t=15]{\includegraphics[scale=0.35]{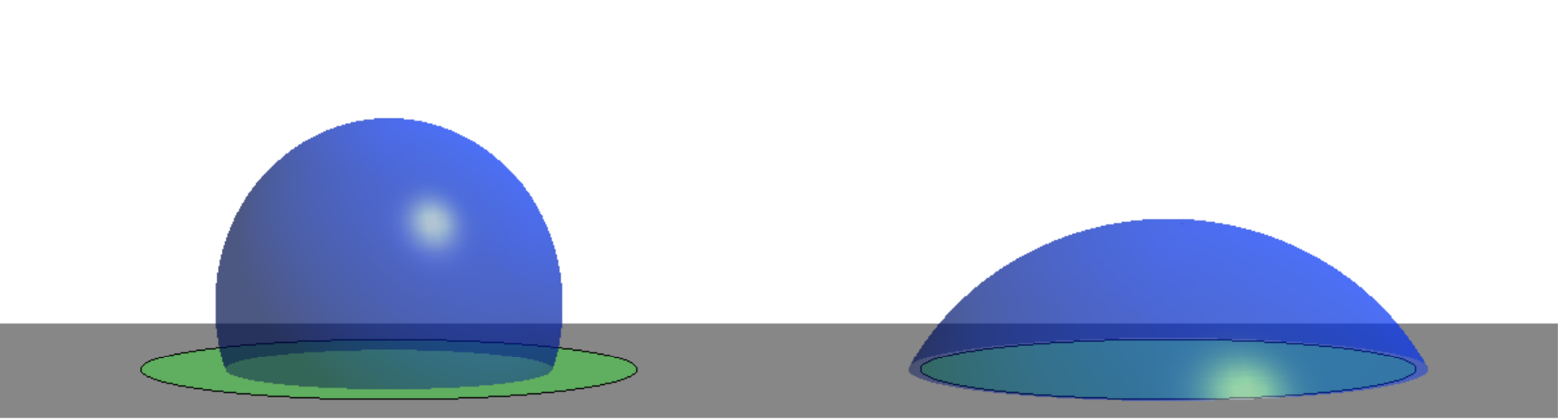}}
     \end{minipage}
\caption{\label{F:patt} Evolution of two drops impacting a patterned surface at (left) $U=4$~m~s${}^{-1}$ and (right) $U=5$~m~s${}^{-1}$.  The hydrophobic surface patterning ($r<1.52$, $\theta_e=110^\circ$) is green whilst the hydrophilic one is grey ($r>1.52$, $\theta_e=60^\circ$).  }
\end{figure}

Consider how topologically different patterns of wettability on a substrate can be used to gain a required level of flow control on a drop once it has been deposited. Here, one such pattern is considered with the aim of changing the final shape of a drop for the same liquid-solid combination by only slightly changing the impact speed, an outcome which is impossible on a homogeneous substrate.  To do so, we pattern an otherwise wettable solid ($\theta_e=60^\circ$) with a circle of nonwettable substrate ($\theta_e=110^\circ$) of radius $1.52$ times that of the drop's initial radius (Figure~\ref{F:patt}).

\begin{figure}\centering
\includegraphics[scale=0.32]{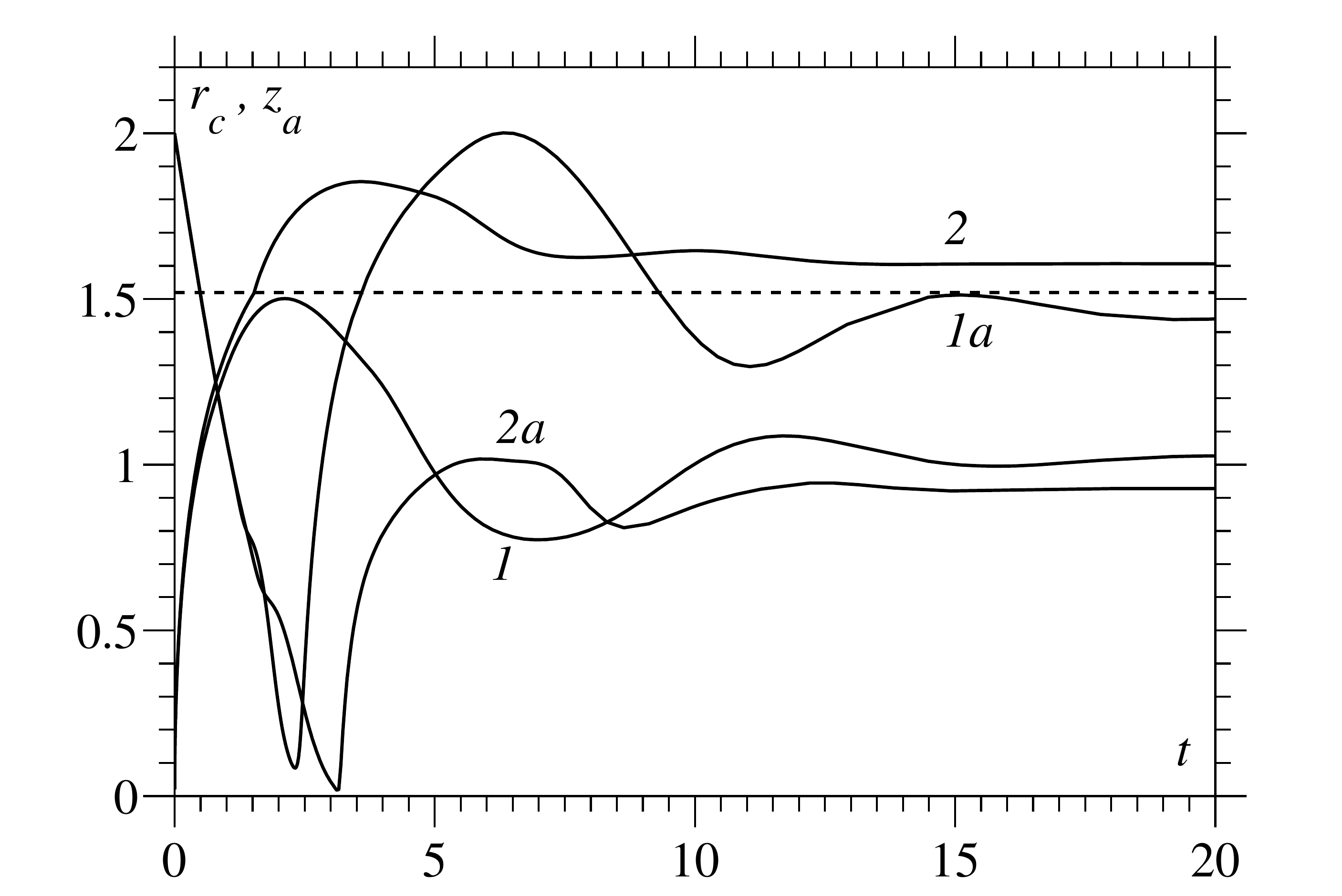}
     \caption{Radius of the contact line (1, 2) and apex height (1a, 2a) as a function of time for two drops impacting a
     patterned surface at 1:~$U=4$~m~s${}^{-1}$ and 2:~$U=5$~m~s${}^{-1}$, respectively.  The boundary between the surface characterized by
     $\theta_e=110^\circ$ $(r<1.52)$ and that defined by $\theta_e=60^\circ$ $(r>1.52)$ is marked with a dashed line.}\label{F:patt_positions}
\end{figure}

From Figure~\ref{F:patt} and Figure~\ref{F:patt_positions}, and the movie of the simulation whose link is here \href{http://youtu.be/AXC1eMHoryE}{Movie}, we see how the desired flow control becomes realizable. On the patterned surface, the equilibrium radius of the area wetted by a drop impacting at $4$~m~s${}^{-1}$ is
found to be $r_c=1.03$, whilst for a $5$~m~s${}^{-1}$ impact it is $r_c=1.61$. This occurs because for the lower
impacting speed the drop is unable to reach the edge of the hydrophobic area to access the more wettable region, and hence behaves as if it is on
a homogeneous surface with wettability defined by $\theta_e=110^\circ$.  For the higher speed of impact the drop is
able to reach the edge of the hydrophobic disc, and encounter the more wettable substrate,
as can be seen by looking at curve 2 in Figure~\ref{F:patt_positions} at $t=2$.  This results in an
increase in the wetting speed and causes the contact line to advance further. This is no guarantee
that the drop's contact line will remain on the more wettable surface as the contact line could
return to the hydrophobic solid, which in turn would enhance the dewetting process.
However, from curve 2 in Figure~\ref{F:patt_positions}, we can see that the contact line's recoil
is relatively shallow, and, in the case considered, it approaches an equilibrium position without ever
encountering the hydrophobic disc again.

%

Thus, it has been shown that custom built substrates can be designed very quickly using our code to determine the specific details, such as the radius of the nonwettable inner circle, that are critical to the success of the product.  Such a computational tool could be used as a precursor to full-scale laboratory experiments and would help to vastly narrow the bounds on potential parameter regimes for a given requirement.  Although we were able to choose different final contact-line radii by varying the impact speed, once the solid had been chosen, we had no control over what this radii would be. With an annulus of hydrophobic surface one would have the possibility of using the impact speed to control the final wetted area. However, the main new physical effect is the one considered here.

\section{Conclusion}

The ability of the computational framework developed in \onlinecite{sprittles11c,sprittles_jcp}, which for the first time models dynamic wetting as an interface formation process, to provide high-accuracy benchmark simulations for flows with large changes in free surface shape has been demonstrated. An initial study into the key physical effects of microdrop impact and spreading, applicable to the inkjet printing regime, has been performed, and the effect of the wettability of a substrate shown to be critical to the drop's dynamics, which can now be recovered in regimes hidden to experimental analysis. We have shown that the strong influence of the substrate's wettability on the drop's dynamics can be utilized to design chemically patterned surfaces allowing one to change considerably the final shape of the drop by only slightly altering the impact speed.  This is an entirely new physical effect which deserves further investigation and fine tuning.

In this article, our focus has been on showing the capabilities of our computational platform and highlighting some of the key physical effects of microdrop impact and spreading.  Future work will be concerned with comparing our results to both existing experimental data in the literature, with the initial simulations showing excellent agreement with those observed in \onlinecite{dong07}, and, in parallel, using our code to design theory-driven experiments.  The latter could identify parameter regimes in which bifurcations in the flow behaviour occur, such as rebound of the drop, dewetting of the centre, etc, and highlight where the differences between the predictions of models proposed in the literature for dynamic wetting will be most prominent.

The finite element framework developed has already been shown to possess reasonable flexibility: it was used to consider flow over patterned surfaces \citep{sprittles07,sprittles09}, to simulate two-phase flow through a capillary \citep{sprittles11c,sprittles_jcp} and here it was used to study the dynamics of free drops and their interaction with solid surfaces of varying types.  This flexibility allows our research programme to branch out and simulate a whole array of different dynamic wetting flows, e.g.\ the coating of fibres \citep{ravinutala02}, where the high speeds of coating in confined areas suggest `hydrodynamic resist' \citep{sprittles_jcp} may be present, or it can be applied to entirely different flows where interface formation has also been shown to be critical, such as the coalescence of liquid bodies, breakup of liquid jets, disintegration of thin films, and other phenomena \citep{shik07}.

\bibliography{main_arxiv}

\end{document}